\documentclass[twocolumn]{aa} 
\usepackage{graphicx}
\usepackage{txfonts}
\usepackage{natbib}
\bibpunct{(}{)}{;}{a}{}{,}
\usepackage{longtable}
\usepackage{aecompl}
\usepackage{amssymb}
\usepackage{amsmath}
\usepackage{multicol}
\usepackage{float}
\usepackage{enumitem}
\usepackage{multirow}
\usepackage{amsfonts}
\usepackage{ccaption}

\newcommand{\krome}{\textsc{krome} }
\newcommand{\kromes}{\textsc{krome}}

\newcommand{\ith}{$i$th }
\newcommand{\jth}{$j$th }

\newcommand{\mC}{\mathrm C}

\newcommand{\mH}{\mathrm H}

\newcommand{\me}{\mathrm e}
\newcommand{\mSi}{\mathrm Si}
\newcommand{\mO}{\mathrm O}

\newcommand{\Hp}{\rm{H}^{+}}

\newcommand{\Hep}{\rm{He}^{+}}
\newcommand{\He}{\rm{He}}

\newcommand{\cp}{\rm{C^{+}}}
\newcommand{\op}{\rm{O}^{+}}

\newcommand{\sip}{\rm{Si^{+}}}
\newcommand{\sipp}{\rm{Si^{++}}}
\newcommand{\expf}[3]{\exp \left(#1\frac{#2}{#3}\right)}
\def\simless{\mathbin{\lower 3pt\hbox
   {$\rlap{\raise 5pt\hbox{$\char'074$}}\mathchar"7218$}}}   
\def\simgreat{\mathbin{\lower 3pt\hbox  
   {$\rlap{\raise 5pt\hbox{$\char'076$}}\mathchar"7218$}}} 
\begin{document}
   \title{A chemical model for the interstellar medium in galaxies}
   \author{S. Bovino 
          \inst{1}
         \fnmsep\thanks{Corresponding author email: stefano.bovino@uni-hamburg.de},
         T. Grassi
          \inst{2},
          Pedro R. Capelo
          \inst{3},
          D. R. G. Schleicher
          \inst{4},
         \and
          R. Banerjee
          \inst{1}
          }

 \institute{$^1$Hamburger Sternwarte, Universit\"at Hamburg, Gojenbergsweg 112, 21029 Hamburg, Germany\\
 $^2$Centre for Star and Planet Formation, Niels Bohr Institute \& Natural History Museum of Denmark, University of Copenhagen, \O ster Voldgade 5-7, DK-1350 Copenhagen, Denmark\\
 $^{3}$Center for Theoretical Astrophysics and Cosmology, Institute for Computational Science, University of Zurich, Winterthurerstrasse 190, CH-8057 Z\"urich, Switzerland\\
 $^{4}$Departamento de Astronom\'ia, Facultad Ciencias F\'isicas y Matem\'aticas, Universidad de Concepci\'on, Av. Esteban Iturra s/n Barrio Universitario, Casilla 160, Concepci\'on, Chile\\
             }

   \date{Received ; accepted}

 
  \abstract
   {}
   {We present and test chemical models for three-dimensional hydrodynamical simulations of galaxies. 
   We explore the effect of changing key parameters such as metallicity, radiation and non-equilibrium versus equilibrium metal cooling approximations on the transition between the gas phases in the interstellar medium.}
   {The microphysics is modelled by employing the public chemistry package \krome  and the chemical networks have been tested to work in a wide range of densities and temperatures. We describe a simple H/He network following the formation of H$_2$, and a more sophisticated network which includes metals. Photochemistry, thermal processes, and different prescriptions for the H$_2$ catalysis on dust are presented and tested within a one-zone framework. The resulting network is made publicly available on the \krome webpage.}
   {We find that employing an accurate treatment of the dust-related processes induces a faster HI--H$_2$ transition. In addition, we show when the equilibrium assumption for metal cooling holds, and how a non-equilibrium approach affects the thermal evolution of the gas and the HII--HI transition.}
   {These  models can be employed in any hydrodynamical code via an interface to \krome and can be applied to different problems including isolated galaxies, cosmological simulations of galaxy formation and evolution, supernova explosions in molecular clouds, and the modelling of star-forming regions. The metal network can be used for a comparison with observational data of CII 158 $\mu$m emission both for high-redshift as well as for local galaxies.}

   \keywords{Astrochemistry --
                Molecular processes -- ISM: evolution, galaxies --
                Methods: numerical
               }

   \authorrunning{Bovino et al.}
   \titlerunning{A chemical model for the ISM in galaxies}
   \maketitle
%

\section{Introduction}
To follow the formation and evolution of galaxies is a computational challenge because many non-linear processes have to be taken into account. For instance, magnetic fields, turbulence, heating and cooling, star formation, feedback, and supermassive black holes, are some of the most important ingredients that make the galactic environment diverse and very complex. Furthermore, the involved spatial and temporal scales span over a wide range, and  it is difficult to explore the dynamics and the physics on all the relevant scales without introducing subgrid-scale models. These are mainly oriented at describing small-scale phenomena such as star formation and stellar feedback \citep{McKee77,Offner2009,Braun2012,Federrath2014,Hopkins2014,Semenov2015} and properties of turbulence \citep{Schmidt2011,Grete15}, and are based on observational constraints. In particular the star formation process, i.e. the conversion of cold gas into stars, is typically tuned according to the Kennicutt-Schmidt (KS) relation \citep{KS1959,KS1963,KS1989,KS1998}, which connects the star formation rate (SFR) to the total gas surface density $\Sigma_{\mathrm{gas}}$. 
Since the work by \citet{Wong2002}, a tight linear correlation between SFR and the H$_2$ content ($\Sigma_{\mathrm{H_2}}$) has been indicated, and it has been shown that the traditional KS relation can break down in the HI regions of  galaxies \citep{Bigiel2008,Leroy2008,Bigiel2010,Schruba2011}. 

The lack of an electric dipole moment makes the H$_2$ line transitions very weak and difficult to be observed, hence CO has  been widely used as a good tracer of the cold phase (see \citealt{Carilli2013} for a recent review). However,  \citet{Krumholz2012} showed that, at metallicities below a few per cent of solar metallicity, star formation could take place in HI-dominated regions, and the argument which correlates the star formation to molecular gas breaks down. In addition, a universal \mbox{turbulence-regulated} star formation law has also been proposed by different authors \citep[e.g.][]{Krumholz2012ApJ,Federrath2013,Salim2015}. It is moreover important to highlight recent observations of the bright CII 158 $\mu$m transition both in high-redshift galaxies with \textsc{alma} \citep{Maiolino2015} and in quasars with \textsc{iram} \citep{Ba2015}, as well as in local galaxies with \textsc{herschel} (e.g. \citealt{Pineda2014}), which provide an alternative path to measure the SFR in the different gas phases of the galaxy. It is then very important to follow the evolution of the gas between the different phases and to include a proper chemical model which can track the time-dependent H$_2$ and/or CII evolution in simulations which aim to study the correlation between SFR and the different gas components.

Analytical and semi-analytical models, some of which are based on equilibrium chemistry, have been proposed to track the formation of molecular hydrogen in three-dimensional (3D) hydrodynamic simulations of galaxies \citep{Kruhmolz2008,Kuhlen2012,Braun2012,Sommerville2015}. A  model which follows the non-equilibrium H$_2$ evolution, coupled with a radiative transfer algorithm, and a simple prescription for metal-line cooling, has been reported by \citet{Gnedin2009}, who also linked the SFR directly to the H$_2$ density, performing simulations with the Adaptive Refinement Tree (ART) code. They included the most important paths for H$_2$ formation and destruction, i.e. photodissociation by UV Lyman-Werner photons, formation on dust, and gas phase formation via H$_2^+$ and H$^-$, and employed a clumping factor $C_{\rho}$ to take into account the unresolved high density regions on small scales. 
This network was extended to  a more detailed model which includes He-based species in a follow-up work \citep{Gnedin2011}, and employed in realistic cosmological simulations of a galaxy at redshift $z = 4$. 

The minimal chemical model by \citet{Gnedin2009} has been implemented in cosmological simulations of galaxies performed with the hydrodynamic codes \textsc{gasoline} \citep{Christensen2012}, and \textsc{ramses} \citep{Tomassetti2015}, which also directly connected the SFR to the H$_2$ content.
\citet{Pelupessy2009} additionally explored an H$_2$-regulated star formation recipe performing cosmological simulations with a time-dependent approach to follow the HI-H$_2$ transition.
However, whether or not the SFR should be directly proportional to the H$_2$ content ($\rho_{\mathrm{H_2}}$) or to the total gas density ($\rho_{\mathrm{gas}}$) is still a matter of ongoing debate, and more observational constraints are needed.

In a recent paper \citet{Hu2015} discussed the connection of star formation to the HI and H$_2$ gas phases, and how this is affected by changing the UV radiation strength, employing a non-equilibrium model for the chemistry. Their chemical network is based on six species, of which only three are followed time-dependently (H$_2$, H$^+$, and CO), and includes the most important heating/cooling processes. The results from high-resolution hydrodynamical simulations for an isolated dwarf galaxy performed with the smoothed-particle hydrodynamics code \mbox{\textsc{gadget-3}}, showed that the dust-to-gas ratio and the radiation background mostly affect the H$_2$ formation rate instead of the SFR, and that the cold phase is dominated by HI.

\citet{Richings2014} presented a complex CO chemical network and performed hydrodynamical simulations of an isolated galaxy assuming a static potential for the dark matter \citep{Richings2015}. Many ingredients, including cosmic-rays and X-rays have been included, and results for different metallicities and UV background strength  have been discussed. 

More recently, a large effort has been made to settle a code comparison of different hydrodynamics codes on galaxy simulations, which share common physics and analysis tools \citep{Kim2014}. However, the \textsc{agora} project mainly aims at studying the atomic gas phase, and to understand if the differences produced by the different hydrodynamic codes can be attributed to the employed physics or the different numerical techniques used to solve the hydrodynamical equations.


In this paper we present a set of chemical models, publicly released via the \krome package\footnote{\url{www.kromepackage.org}}, that can be used to study the transitions between the different gas phases in galaxy simulations with different degrees of accuracy. We provide a method for non-equilibrium H$_2$ prescription, and a more complex network including metals which can be employed to follow the evolution of CII at different metallicities. The paper is organized as follows. In Section \ref{sect:micro}, we present the most important physical ingredients which regulate the thermal and chemical evolution of the gas, and their computational implementation in the public code \kromes. In Section \ref{sec:models_test}, we test the chemical networks and the microphysics employing a simple one-zone collapse framework. We explore four different models under different conditions, i.e. changing the metallicity, the chemical initial conditions, the radiation background strength, and the dust-related physics. We finally provide our conclusions in Section \ref{sec:conclusions}.

\section{Microphysics}\label{sect:micro}
\subsection{The KROME package}
The microphysics and the chemical networks described in the following sections have been developed within the publicly available astrochemistry package \krome \citep{Grassi2014}. It consists of a python module which, given a chemical network and a series of physics-related options, generates the code to solve a system of coupled ordinary differential equations (ODEs) which follow the thermo-chemical evolution of the gas. \krome includes pre-compiled networks and different modules for the microphysics, as for instance  cooling/heating functions, photochemistry, and a dust module for cooling and molecule catalysis. Since its release, \krome has been employed to study a variety of astrophysical problems, e.g. the formation of supermassive black holes under UV radiation and in the presence of dust \citep{Latif2014MNRAS,Latif2015,Latif2015MNRAS}, the collapse of low-metallicity minihaloes \citep{Bovino2014}, formation of primordial stars \citep{Bovino2014MNRAS},  the chemical evolution of self-gravitating primordial disks \citep{Schleicher2015}, the formation of the first galaxies \citep{Prieto2015}, the formation of very massive stars \citep{Katz2015}, and star-forming filaments in molecular clouds \citep{Seifried2015}. A wide range of physical and chemical conditions has  been explored by employing \krome with different hydrodynamical codes (e.g. \textsc{enzo}, \textsc{ramses}, and \textsc{flash}) showing its flexibility. In addition, \krome employs the accurate and stable chemical solver  \textsc{dlsodes} \citep{Hindmarsh83}, which is suitable for studying highly non-linear problems where the spatial and temporal domains vary by several orders of magnitude. For more details about \krome we refer to the code paper \citep{Grassi2014} and  to the website \url{http://www.kromepackage.org}.

\begin{figure*}
	\centering
	\includegraphics[scale=0.4]{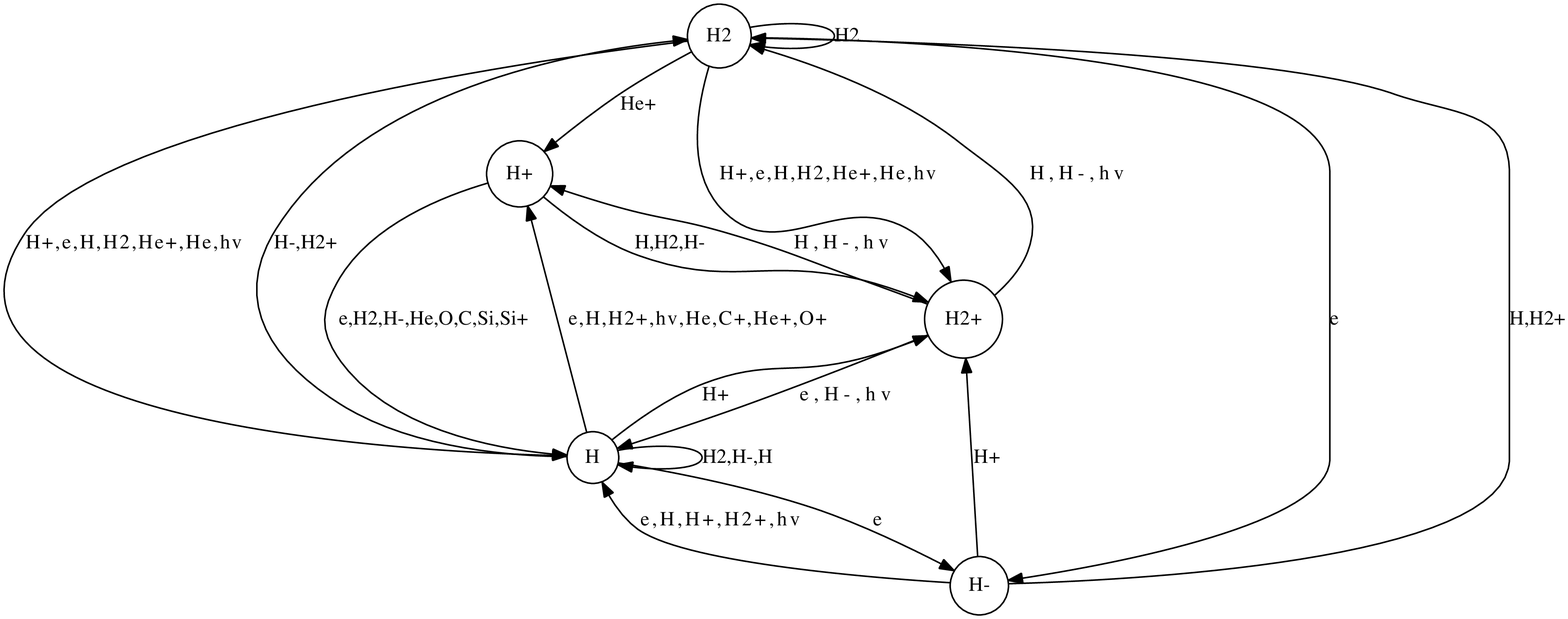}\hspace{1cm}
	\includegraphics[scale=0.4]{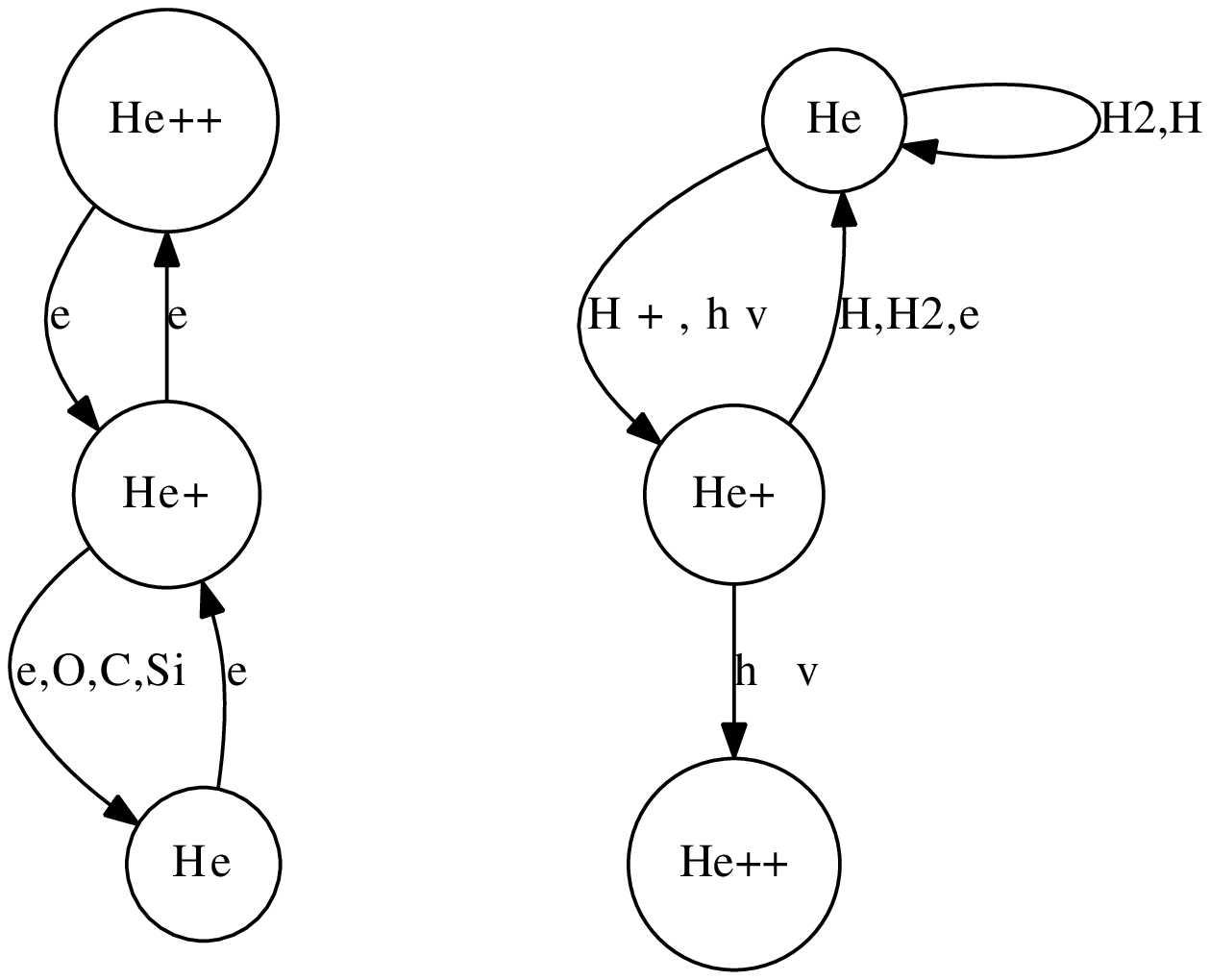}
	\includegraphics[scale=0.4]{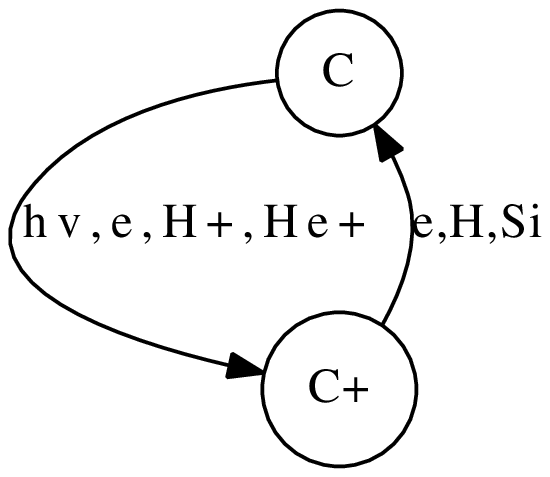}
	\includegraphics[scale=0.4]{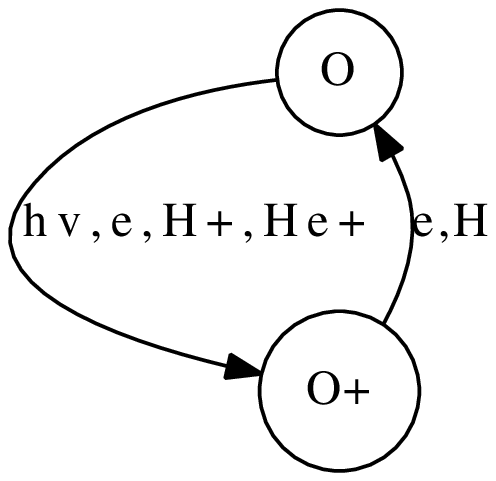}
	\includegraphics[scale=0.4]{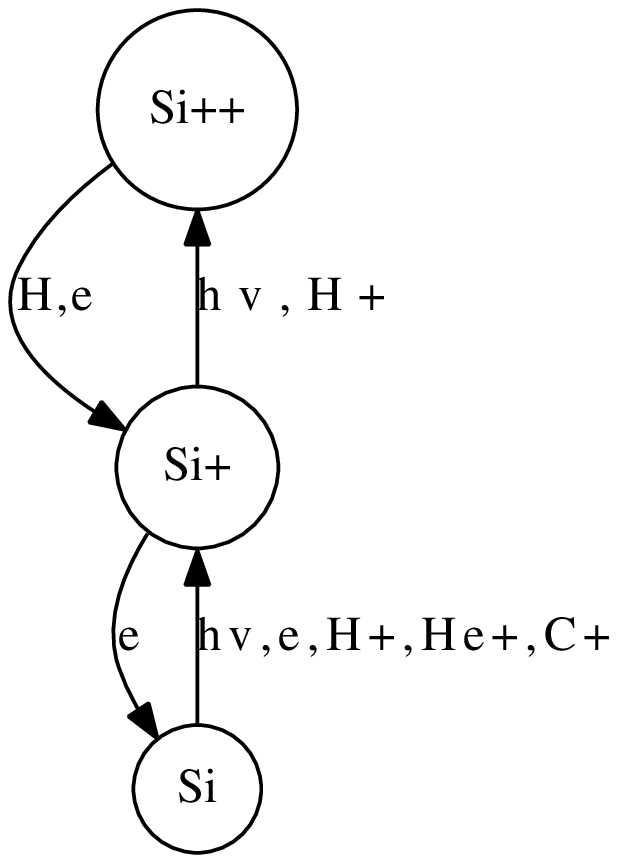}
	\caption{Network graphs for H, He, C, Si, and O as the main hubs. Note that helium has been split into two small sub-networks, the one on the left showing the reactions with metals, and the one on the right the reactions with  hydrogen-based species and photons. The network graphs can be obtained by using the \krome tool \textit{pathway.py}.}\label{fig:graph1}
\end{figure*}

\subsection{Chemical models}
In this Section we present different chemical models which can be employed in simulations of galaxy formation and  evolution, together with the main physical ingredients which regulate the thermal evolution of the gas and  its dynamics through the process of star formation. The models have been tested to work for a wide range of temperatures, namely from 10 to 10$^9$~K. 
The single ingredients presented in this paper are all implemented in  \kromes, and can be easily added/removed to create models of different complexity. Here we present models with increasing level of complexity as shown in Table \ref{tab:models}.


\begin{table*}
\centering
\caption{List of included thermal processes.}
\centering
\begin{tabular}{llllc}
	\hline
	Model \# & Network & Metal cooling  & Dust & Photochemistry \\
	\hline
	Model I & H/He & tables (10--10$^9$ K) & approx. H$_2$ formation$^a$& - \\
	Model Ia & H/He & tables (10--10$^9$ K) & approx. H$_2$ formation$^a$& \checkmark \\
	Model II & H/He & tables (10--10$^9$ K) & H$_2$  (size-dep.)$^b$& - \\
	Model IIa & H/He & tables (10--10$^9$ K) & H$_2$ + cooling  (size-dep.)$^b$& - \\
	Model III & H/He & tables (10--10$^9$ K) & H$_2$ + cooling (size-dep.)$^b$& \checkmark \\
	Model IV & H/He + metals & non eq. (10--10$^4$ K)  + tables (10$^4$--10$^9$ K)& H$_2$ + cooling (size-dep.)$^b$ & \checkmark \\
\hline
\vspace{-0.3 in}
\label{tab:models}
\end{tabular}
\vspace{0.2cm}
\\$^a$ we refer here to the rate reported by \citet{Gnedin2009,Christensen2012} and discussed in Section \ref{sec:h2formation}.\\
$^b$ rates from \citet{Cazaux2009} and cooling from \citet{Hollenbach1979}, see discussion in Section. \ref{sec:h2formation}.\\
\end{table*}


In the following  we will refer to the total number density (cm$^{-3}$) as $n_{\mathrm{tot}}$, to the  atomic neutral hydrogen (HI) number density as $n_{\mathrm{H}}$, while the total density of hydrogen nuclei is denoted as

\begin{equation}
n_{\mathrm{H}_{\mathrm{tot}}} = n_{\mathrm{H}} +  n_{\mathrm{H^+}} + n_{\mathrm{H^-}} + 2n_{\mathrm{H_2^+}} + 2n_{\mathrm{H_2}}. 
\end{equation}
The number density of the single species $j$ is given as $n_j$. 



\subsubsection{Reaction rates}
The first three models include a selection of the most important reactions which contribute to the formation/destruction of molecular hydrogen in a gas consisting of a mixture of H- and He-based species. It includes H, He, He$^+$, He$^{++}$, H$^+$, H$^-$, H$_2$, H$_2^+$, and electrons. The aim of these models is to provide a basic/minimal chemical network which is suitable to follow the transition between the atomic and molecular gas phase, but keeps the computational cost at a reasonable level to be employed in 3D simulations of galaxy evolution. For users who are interested in a more detailed chemical model, a network including several metal species is outlined below and employed in Section \ref{sec:metal_noneqCool}. The list of reactions is reported in Table \ref{tab:rates} together with references and is split into hydrogen/helium species and metals (C, C$^+$, Si, Si$^+$, Si$^{++}$, O, O$^+$). We refer to the complete network as \textit{react\_galaxy}, downloadable with the \krome package\footnote{\url{https://bitbucket.org/tgrassi/krome}}. The rates for the metal network are adopted from \citet{Glover2007} and are the same as employed in \citet{Bovino2014}. Three-body reactions are not included due to the low-density regime involved in the physical problem considered here (with number densities up to 10$^4$--10$^5$ cm$^{-3}$).

In Fig.~\ref{fig:graph1}, we show a sketch of the complete chemical network, where we distinguish hydrogen, helium, and metal sub-networks with the most important paths to form and destroy the different species. As it can be seen from Fig. \ref{fig:graph1}, this is a H$_2$-oriented network including important reactions involving metals and helium. We have a total of 16 species and 74 reactions.

\subsubsection{Photochemistry}
The presence of a radiation background alters the chemical evolution by ionising and dissociating atoms and molecules. Here we consider the photo-processes reported in Table \ref{tab:photo} due to a UVB/X-ray radiation background from quasars and galaxies \citep{Haardt2012} up to 100 keV. The cross-sections, where possible, have been taken from standard databases: the SWRI database\footnote{\url{http://phidrates.space.swri.edu}} and the Leiden database\footnote{\url{http://home.strw.leidenuniv.nl/~ewine/photo/}}. \krome also includes an internal database based on the work by \citet{Verner1996} and \citet{Verner1996_2}. The frequency (energy) domain is divided by \krome in $N$ bins and the photo-rates ($k_{ph}$) and the photo-heating rates ($\Gamma_{ph}$) are computed by employing the following discrete formulae,

\begin{eqnarray}
	k_{ph}& =&\frac{4\pi}{h}\sum_{i=1}^{N}\frac{J_i(\langle E_i\rangle)\sigma_i(\langle E_i\rangle)}{\langle E_i\rangle}(E_i^R - E_i^L),\\
	H_{ph}&=&\frac{4\pi}{h}\sum_{i=1}^{N}\frac{J_i(\langle E_i\rangle)\sigma_i(\langle E_i\rangle)}{\langle E_i\rangle}(\langle E_i\rangle - E_0)(E_i^R-E_i^L),
\end{eqnarray}

\noindent where $h$ is the Planck constant, $J_i$ is the radiation background in units of eV cm$^{-2}$ sr$^{-1}$ Hz$^{-1}$ s$^{-1}$, and $\sigma_i$ and $E_0$ are, respectively, the cross-section (in cm$^2$) and the threshold energy (in eV) of the given photo-process. $\langle E_i\rangle$~=~$(E^R+E^L)/2$ is the average photon energy of a single bin of size $\Delta E$~=~$E^R - E^L$, with $R$ and $L$ standing for \textsc{right} and \textsc{left}. The final photo-rate is in units of s$^{-1}$. 

The photo-heating rate (in eV s$^{-1}$) is then used to compute the final heating $\Gamma_{ph}^j = H_{ph}^j n_j$ for every \jth photo-process, with $n_j$ being the abundance of the ionised/dissociated species in cm$^{-3}$.

In this work we used 10000 photo-bins to ensure convergence on the final photo-rates\footnote{This is an input parameter for \krome that can be easily customized by the users, as well as the bins'spacing, and their energy position.}. These must be re-computed if the radiation background evolves over  time. We consider an optically thin gas and a time-independent radiation background at redshift $z = 3$. A comparison of this background with other standard radiation fields is discussed in Section \ref{sec:metalparam}. The effect of reducing the incident radiation and adding shielding is also reported in the same Section. We do not discuss here the photochemistry induced by galactic UV photons as this is strongly connected to the way the radiation is treated in hydrodynamical codes. For instance, the UV radiation should partly come from stellar sources within the galaxy and requires a proper modelling of the radiative transfer. For more details on how to handle UV sources within simulations of galaxy evolution we refer to previous work \citep{Gnedin2009,Gnedin2011,Christensen2012,Tomassetti2015}. However, \krome can handle any kind of radiation, hence our discussion on how to compute the photo-rates does not change. 

Note that \krome has been designed to be coupled with multi-frequency radiative transfer codes, and it provides a flexible interface that allows to set any bin-based discretization of the impinging radiation flux.


\begin{table*}
        \caption{List of photo-processes included in our chemical network.}\label{tab:photo}
        \begin{tabular}{@{}lllc}
                \hline\hline
                Reaction & Cross-section (cm$^2$) & & Ref.\\
                \hline
  P1\dots\dots ~~H + $\gamma$ $\rightarrow$ H$^+$ + e$^-$  & \krome database   & & 1 \\
  P2\dots\dots ~~He + $\gamma$ $\rightarrow$ He$^+$ + e$^-$  & \krome database   & & 1 \\
    P3\dots\dots ~~He$^+$ + $\gamma$ $\rightarrow$ He$^{++}$ + e$^-$  & \krome database   & & 1 \\
    P4\dots\dots ~~H$^-$ + $\gamma$ $\rightarrow$ H + e$^-$  & SWRI database   & & 2 \\
    P5\dots\dots ~~H$_2$ + $\gamma$ $\rightarrow$ H$_2^+$ + e$^-$  & SWRI database   & & 2 \\
    P6\dots\dots ~~H$_2^+$ + $\gamma$ $\rightarrow$ H$^+$ + H  & LEIDEN database   & & 3\\
    P7\dots\dots ~~H$_2^+$ + $\gamma$ $\rightarrow$ H$^+$ + H$^+$ + e$^-$  &   dex$[-16.926-4.528\times 10^{-2} E+2.238\times 10^{-4} E^2+4.245\times10^{-7} E^3]$  & $30 < E < 90$$^a$ & 4\\
    P8\dots\dots ~~H$_2$ + $\gamma$ $\rightarrow$ H + H  & Direct &  & 5\\
     P9\dots\dots ~~H$_2$ + $\gamma$ $\rightarrow$ H$_2^*$ $\rightarrow$ H + H  & Solomon &  & 6\\
    \hline
    \hline
  	Metals &&&\\
    \hline
     P10\dots\dots ~~C + $\gamma$ $\rightarrow$ C$^+$ + e$^-$  & \krome database  &  & 1\\
     P11\dots\dots ~~O + $\gamma$ $\rightarrow$ O$^+$ + e$^-$  & \krome database  &  & 1\\
     P12\dots\dots ~~Si + $\gamma$ $\rightarrow$ Si$^+$ + e$^-$  & \krome database  &  & 1\\
  P13\dots\dots ~~Si$^+$ + $\gamma$ $\rightarrow$ Si$^{++}$ + e$^-$  & \krome database  &  & 1\\     
     \hline
\end{tabular}
1:\citet{Verner1996,Verner1996_2}, 2: \url{http://phidrates.space.swri.edu}\\ 3: \url{http://home.strw.leidenuniv.nl/~ewine/photo/}
4: \citet{Shapiro1987}\\ 5: \citet{Abel97} 6: this work, based on the formula reported by \citet{Glover2007}\\
$^a$ $E$ is the energy in eV
\end{table*}

\subsubsection{Photodissociation of H$_2$}\label{sec:photoch}
In the presence of an ionising radiation background there are two processes which can photodissociate molecular hydrogen: (i) excitation to the vibrational continuum of an excited electronic state (P8 in Table \ref{tab:photo}), and (ii) the two-step Solomon process (P9 in Table \ref{tab:photo}). These processes have been proposed and widely discussed in \citet{Stecher1967,Allison1969,Abel97,Glover2001,Gay2012}, and we refer to those papers for additional details. The two processes have different thresholds: the ``direct'' dissociation occurs for energies above 14.16~eV for ortho-hydrogen and 14.68~eV for para-hydrogen, and it is important in the presence of a strong ionising UV flux (e.g. in HII regions). The Solomon process occurs in the Lyman (B$^1\Sigma_u^+$) and Werner (C$^1\Pi_u$) bands in a very tiny energy window (11.25--13.51 eV). For the ``direct" process we employ the fit proposed by \citet{Abel97} based on the data by \citet{Allison1969}, whereas the two-step Solomon process we employ the rate suggested by \citet{Glover2007}, given by

\begin{equation}
	k_{ph}(\mathrm{H_2}) = 1.38\times 10^9 ~~ \mathrm{s^{-1}}~ \frac{J(h\overline{\nu} = 12.87~ \mathrm{eV})}{\mathrm{erg ~ s^{-1} cm^{-2} Hz^{-1} sr^{-1}}}.
\end{equation}

\subsection{Thermal processes}\label{sec:thermal}
The thermal state of the gas is regulated by many processes which we report in Table \ref{tab:thermal} and discuss in the following sections. 
Due to resolution limits in  hydrodynamical simulations, it is sometimes necessary to impose a temperature floor, which might also depend on the density. We apply it to the total cooling $\Lambda(T)$ in the following way,

\begin{equation}
	\Lambda_{\mathrm{eff}} (T) = \Lambda (T) - \Lambda (T_{\mathrm{floor}}),\\
\end{equation}
as for example in \citet{Safranek2014}.
\noindent Even though the range of temperatures over which our modelling of thermal processes is valid is quite wide ($10-10^9$~K, see Table~\ref{tab:models}), in most of the tests discussed in this work we assume a \mbox{density-independent} $T$$_{\rm floor} = 100$~K, since most 3D hydrodynamical simulations of galaxy evolution, out of resolution concerns, cannot properly model colder gas. We remind the reader, however, that $T_{\rm floor}$ can be changed by the user, according to the problem and resources one has.

\subsubsection{Atomic and molecular cooling}
The standard atomic cooling (atom recombinations, collisional ionisations, excitations, and Bremsstrahlung of ions) and the Compton cooling from the cosmic microwave background (CMB) have been adopted from \citet{cen92}. H$_2$ roto-vibrational cooling is an update to \citet{Glover2008} which includes the new rates reported by \citet{Glover2015}. 

Following \citet{Hollenbach1979} and \citet{Omukai2000}, we include chemical cooling coming from the destruction of H$_2$ in the gas phase  (reactions 18, 19, in Table \ref{tab:rates}). This is discussed  in detail in the \krome paper \citep{Grassi2014}.

Recombination reactions on dust can also contribute to the cooling (see e.g. reactions 29, 30 in Table \ref{tab:rates}); we have included this effect following \citet{BakesTielens94}, as discussed in Section \ref{sec:photoelectricH}.

\subsubsection{Metal cooling in previous work}
Cooling from metal line transitions is one of the most important contributions at low-densities which regulates the thermal evolution of the gas and then the star formation process. The complexity of the interstellar medium (ISM) makes it challenging to evaluate it at run time even when we consider a sub-ensemble of the most important species (C, N, O, Ne, Si, Mg, S, Ca, and Fe). The large amount of ionisation states, transitions and collisional processes (and different colliders) makes in fact the problem computationally prohibitive within hydrodynamical simulations, as we should solve within every chemical time-step a linear system of equations to calculate the level populations for every metal species and then the cooling efficiencies. 
This problem has been extensively discussed in the literature and there has been a lot of effort to produce simplified models or to provide cooling tables which can be easily employed in hydrodynamic simulations. 

Cooling tables for metals have been calculated over the years in different ways and assuming different approximations and they turned out to be the most common approach in simulations of galaxies and of the intergalactic medium. We can distinguish between four different cases: pure collisional ionisation equilibrium (CIE) tables \citep{SutherlandDopita93}, collisional ionisation non-equilibrium (CINe) tables \citep{Gnat2007,Gnat2012},  tables which consider a photo-ionising background in photo-ionisation equilibrium (PIE) as in \citet{Wiersma2009}, and photo-ionisation non-equilibrium (PINe) tables, as the recent tables reported by \citet{Vasiliev2011} and \citet{Oppenheimer2013}. \citet{Richings2014} extended these tables to a larger range of densities and improved the model below 10$^4$ K for C and O.

Most of these tables have been calculated with the photo-ionisation code \textsc{cloudy} \citep{Ferland98}. Usually \textsc{cloudy} provides equilibrium cooling tables (CIE or PIE), which are evaluated at a  fixed temperature, density, and metallicity ($Z/Z_\odot$) evolving the system to reach the equilibrium (when ionisations balance recombinations). In the approach by \citet{Gnat2007}, \citet{Vasiliev2011}, and \citet{Oppenheimer2013}, who explored the effect of non-equilibrium chemistry on the final cooling functions, \textsc{cloudy} was mainly used to evaluate the cooling functions, requiring as input the non-equilibrium ionisation fractions of the metal ions-species evaluated by solving a system of ODEs over the time, for a given temperature and density. 

\citet{Gnat2007} computed the cooling without any ionisation background, i.e. in collisional ionisation non-equilibrium and provided tables for the non-equilibrium ion fractions $x_{ion}$ and ion-by-ion cooling efficiencies $\Lambda_{e,ion}$. The sum of ionic cooling efficiencies, weighted by the non-equilibrium ion densities, then provides an efficient way to compute the total cooling of a given metal. They reported clear examples on how to use the cooling tables in \citet{Gnat2012}. The final total cooling rate is written as $\Lambda_{tot} = n_e\sum_j (n_j \sum_{ion} x_{j,ion}\Lambda_{e,ion})$, with $n_j$ being the total amount of the considered coolant metal, and the $ion$ subscript runs over the ionisation states of the \jth metal.

\begin{table}
\centering
\begin{minipage}{84mm}
\caption{List of included thermal processes.}
\centering
\begin{tabular}{lr}
\hline 
Process & References\footnote{1 - \citet{cen92}; 2 - \citet{BakesTielens94}; 3 - \citet{Shen2010}; 4 - this work, from data by \citet{Glover2008,Glover2015};  5 - \citet{Omukai2000}; 6 - \citet{Grassi2014}; 7 - \citet{BakesTielens94}; 8 - \citet{Burton1990}; 9 - \citet{Hollenbach1979,Omukai2000}; 10 - \citet{Hollenbach1979}} \\
\hline
\multicolumn{2}{c}{\textbf{Cooling Processes}} \\ 
\hline
H, He, He$^+$ excitations & 1 \\
H, He, He$^{+}$ collisional ionisations & 1 \\
H$^{+}$, He$^{+}$, He$^{++}$ recombinations & 1 \\
Compton cooling from the CMB & 1 \\
Bremsstrahlung & 1 \\
Grain-surface recombination & 2 \\
Metal line cooling & 3,6 \\
H$_{2}$ roto-vibrational cooling & 4 \\
H$_{2}$ collisional dissociation & 5 \\
Grain-gas thermal exchange & 9\\
\hline
\multicolumn{2}{c}{\textbf{Heating Processes}} \\ 
\hline
Photo-heating & 6 \\
Dust photoelectric effect & 7 \\
H$_{2}$ UV pumping & 8 \\
H$_{2}$ formation; gas phase & 9 \\
H$_{2}$ formation; dust grains &  10 \\
\hline
\vspace{-0.3 in}
\label{tab:thermal}
\end{tabular}
\end{minipage}
\end{table}

\citet{Vasiliev2011} provided cooling tables for the gas exposed to different radiation backgrounds evaluating the non-equilibrium ionisation fractions by adopting a similar approach as in \citet{Gnat2007}. 
They explored low density gas and different metallicites going from 10$^{-3}$ to 1~$Z_\odot$ and showed  that assuming PIE leads to an overestimate of the cooling for $T > 10^6$ K, as the gas in non-equilibrium keeps overionised. At higher densities, collisions dominate over photo-ionisations and this effect is reduced. An increase on the radiation strength further decreases the final cooling rate. 

The recent work by \citet{Oppenheimer2013} provided an additional overview on the effect of the radiation and non-equilibrium chemistry on the metal cooling efficiencies. Their approach works better for $T > 10^4$ K as they assume the collisions to be dominated by electrons, but if radiation is included they can consider the results below $T < 10^4$ K as reasonable.
They explored the four approaches discussed above, namely the CIE, CINe, PIE and PINe, and compared the results focussing mostly on the regime of temperatures above 10$^4$ K.

An important final statement of their work is that the presence of a radiation background suppresses cooling either in equilibrium or non-equilibrium. 

\begin{figure}
	\includegraphics[scale=0.7]{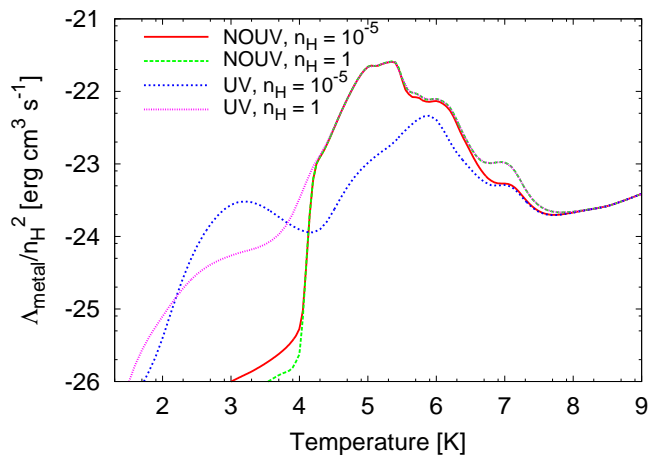}
	\caption{Metal line cooling from equilibrium (PIE) tables for a gas of metallicity $Z = 0.5~Z_\odot$ at redshift $z = 3$. The case without a radiation background (NOUV) and with a radiation background (UV) are both shown for two different densities: $n = 10^{-5}$ and $n = 1$ cm$^{-3}$.}\label{fig:metal}
\end{figure}

\subsubsection{Metal cooling in \krome}
For the metal cooling at $T \geq 10^4$ K we employ an updated version of the PIE tables presented by \citet{Shen2010}, which have been computed with \textsc{cloudy 10.1} in the presence of the extragalactic radiation by \citet{Haardt2012}. The tables are valid in a range of temperatures between 10 $\le T \le 10^9$ K, density $10^{-9} \le n_{\mathrm{H}_{\mathrm{tot}}} \le 10^4$ cm$^{-3}$, and redshift $0 \le z \le 15.1$ and scale linearly with the metallicity:

\begin{equation}\label{eq:metaleq}
	\Lambda_{\mathrm{metal}}^{eq}(T,n_{\mathrm{H}_\mathrm{tot}},z,Z) = \Lambda_{\mathrm{metal},\odot}^{eq} (T,n_{\mathrm{H}_\mathrm{tot}},z) ~Z/Z_\odot.
\end{equation}
The equilibrium cooling computed by \citet{Shen2010} includes all metals up to an atomic number of 30 (Zn). The H and He contributions are not included as they are computed time-dependently based on the non-equilibrium chemical evolution. The heating rate coming from the photo-ionisation of metals is also included in the same fashion. 
In Fig. \ref{fig:metal} we show the equilibrium metal cooling function for densities of 10$^{-5}$~cm$^{-3}$ and 1 cm$^{-3}$ and at redshift $z$~=~3, for the case with and without radiation at a metallicity $Z = 0.5~Z_\odot$. Those cooling functions are comparable to the former work by \citet{Shen2010}, where they also included the contribution coming from H and He that we do not show here.

\krome includes the machinery to solve the linear system for the individual metal excitation levels time-dependently for temperatures below 10$^4$ K, and provides the non-equilibrium metal cooling for the most important atoms/ions\footnote{Note that the approach to evaluate the metal line cooling has general validity but, due to the limited validity range of the collisional rates, we can only apply it below 10$^4$ K.}. This method has been tested in a recent study by \citet{Bovino2014}, and includes the non-equilibrium cooling of C, C$^+$, Si, Si$^+$, O, O$^+$, Fe, and Fe$^+$, as described in \citet{Grassi2014} based on  \citet{Glover2007}, and \citet{Maio07}. \citet{Richings2014} have shown that under the presence of a strong photo-ionising background as the one employed in this work, the most important coolants at $T < 10^4$ K are C$^+$, Si$^+$, and Fe$^+$, together with hydrogen. In general at low-temperatures or  higher densities the excess of electrons favours forbidden line transitions of low-ionised metals such as C, C$^+$, Si, Si$^+$, and O, as also reported by \citet{Shen2010} and \citet{Wolfire2003}. Our non-equilibrium approach below 10$^4$ K is  able to capture the most important metal contributions to the total cooling function in an accurate way, i.e. solving the non-equilibrium system at run time according to the network we employ in our simulations. 

It is worth noting that this approach is different from the non-equilibrium tables computed for example by \citet{Vasiliev2011}, and \citet{Oppenheimer2013} (PINe approach), which depend not only on the initial conditions (i.e. ionic composition and temperature) employed to solve the system of ODEs \citep{Vasiliev2010}, but are also bound to a fixed chemical network which in most cases does not include molecules. Our cooling is instead accurately evaluated on the fly according to the evolution of the given chemical network (that can change), and is fully consistent with the conditions explored in the physical problem users want to pursue.


The final metal cooling in the  temperature  range of \mbox{10--10$^9$}~K is then expressed as a combination of equilibrium and non-equilibrium cooling,

\begin{equation}\label{eq:metaltot}
		\Lambda_{\mathrm{metal}} = f_1 \Lambda_{\mathrm{metal}}^{eq}  + f_2 \Lambda_{\mathrm{metal}}^{non-eq}, 
\end{equation}
where $f_1$ and f$_2$ are two  functions which allow a smooth transition between equilibrium and non-equilibrium cooling around \mbox{$T$ = 10$^4$}~K and are defined as

\begin{eqnarray}
	f_1(T) &=&\frac{1}{2}\{\tanh[c_{\mathrm{sm}}~(T-10^4)]+1.0\},\\
	f_2(T) &=&\frac{1}{2}\{\tanh[c_{\mathrm{sm}}~(-T+10^4)]+1.0\},
\end{eqnarray}
with $c_{\mathrm{sm}} = 10^{-3}$ and $f_1(T) + f_2(T) = 1$, for every $T$. The smoothing factor $c_{\mathrm{sm}}$ is necessary to help the solver in reaching the convergence.

Given the temperature range of validity for the equilibrium table one can choose between  a simple approach (Eq.~\ref{eq:metaleq}) over the whole range or a more sophisticated treatment (Eq.~\ref{eq:metaltot}). 
The aim of this work is in fact to release a chemical model within the publicly available package \kromes, which can be easily modified by the users to allow a lower or higher degree of complexity in large scale simulations.

%

\subsubsection{Dust cooling}
The dust cooling is calculated based on \cite{Hollenbach1979} who employ the following expression,
\begin{equation}\label{eq:dustcool}
        \Lambda_\mathrm{dust} =  2 k_B \pi a^2  n_{tot}n_d v_g  (T-T_d)\,,
\end{equation}
where $n_{tot}$ and $n_d$ are the total gas number density and the grain number density, respectively, $\pi a^2$ is the dust grain effective cross-section, with $a$ being the grain size, $v_g$ is the gas velocity defined as $v_g = \sqrt{8k_BT/\pi/m_p}$, with $m_p$ being the proton mass, and $T_d$ is the dust temperature. In the following, we will for simplicity assume a constant dust temperature of 10~K, which is typical for molecular clouds. The user is however free to adopt a different description, for instance calculating the dust temperature from the heating-cooling balance of the grains \citep{Grassi2014}. Note that when the dust is hotter than the gas (i.e. $T_d > T$), equation (\ref{eq:dustcool}) acts as a heating term for the gas.

We tabulate the dust cooling similarly to the H$_2$ formation rate on dust as explained in Section \ref{sec:h2formation}, and we show in Section \ref{sec:model1} that, under the conditions discussed here, its effect is negligible. 

\subsubsection{H$_2$ photo-heating}
The most important contribution to the heating of the gas under the conditions discussed in this paper is the photo-heating. We include the photo-heating of atoms and molecules due to ionisations and photodissociations. The latter also includes the H$_2$ UV pumping and H$_2$ direct photodissociation heating as explained below.

The energy budget produced by the Solomon process is converted into heating through two different paths, the direct dissociation, which provides 0.4 eV as kinetic energy, and the energy released by collisional de-excitation of the excited vibrational levels of the H$_2$ electronic ground state, which is commonly called H$_2$ UV pumping heating. 
The final heating  is the sum of the two contributions and is given by
\begin{equation}
	\Gamma_{\mathrm{tot}} (\mathrm{H_2}) = (H_{pd}^{\mathrm{dir}} + H_{pd}^{\mathrm{pump}}) n_{\mathrm{H_2}},\\
\end{equation}
where
\begin{eqnarray}
	H_{pd}^{\mathrm{dir}} &=& 6.4\times 10^{-13} k_{ph}(\mathrm{H_2})\\
	H_{pd}^{\mathrm{pump}} &=& 3.5\times 10^{-12} (8.5\times k_{ph}(\mathrm{H_2}))\left[1+\frac{n_{cr}}{n_{\mathrm{H}_{\mathrm{tot}}}}\right]^{-1} .
\end{eqnarray}
 The total heating $\Gamma_{\mathrm{tot}} (\mathrm{H_2})$ is expressed in terms of erg cm$^{-3}$ s$^{-1}$.
We have followed \citet{Burton1990}, \citet{Glover2007}, and \citet{Draine1996}, who assume that the pumping rate is 8.5 times larger than the photodissociation rate $k_{ph}(\mathrm{H_2})$. The critical density $n_{cr}$ is the density at which the collisional de-excitations become comparable to the spontaneous radiative processes, i.e. this process only becomes important for $n_{\mathrm{H}_{\mathrm{tot}}} \geq n_{cr}$, which is $\sim$10$^4$ cm$^{-3}$. For the definition of the $n_{cr}$ we refer to \citet{Hollenbach1979} (see also \citealt{Omukai2000}).

\subsubsection{Photoelectric heating by dust grains}\label{sec:photoelectricH}
UV radiation with photon energies of about 6 eV or more can remove electrons from interstellar dust grains (GR), that increase the energy of the surrounding gas by subsequent Coulomb collisions. This heating process is called dust photoelectric effect and is due to the  reactions
\begin{equation}
			\mathrm{GR}^{+(q)} + \gamma \rightarrow  \mathrm{GR}^{+(q + 1)} + \mathrm{e^-},
		\end{equation}
\noindent where $q\ge 0$ is the charge number.
\noindent The above process is often in competition with its inverse, the recombination reaction 
\begin{equation}
\mathrm{GR}^{+(q)} + \mathrm{e^-} \rightarrow  \mathrm{GR}^{+(q - 1)},
	\end{equation}
	which removes energy from the gas producing cooling. The net dust photoelectric heating has been parametrized by \citet{BakesTielens94} who introduced the charging parameter
	
	\begin{equation}
	\psi = \left(\frac{G_0 T^{1/2}}{n_e}\right),
	\end{equation}
	with $G_0$ being the Habing flux and $n_e$ the electron number density in cm$^{-3}$ . We obtain $G_0$ by
	summing the flux over the photo-bins which are in the range between 6--13.6 eV, normalized to the average interstellar radiation field flux
	\begin{equation}
		G_0  = \frac{4\pi}{h}\frac{\sum_{\nu=6-13.6} J(E)\Delta E}{1.6\times 10^{-3} \mathrm{erg~cm^{-2}~s^{-1}}},\\
	\end{equation}
	as reported by \citet{Omukai2008}.
	
\noindent The net photoelectric heating is defined as
	\begin{eqnarray}
		\Gamma_{pe}^{net} &=& \Gamma_{pe}-\Lambda_{rec},\\
		\Gamma_{pe} &=& 1.3\times 10^{-24}\epsilon G_0 n_{\mathrm{H}_{\mathrm{tot}}},\\
		\Lambda_{rec} &=& 4.65\times 10^{-30} T^{0.94} \psi^{ 0.735T^{-0.068}} n_e n_{\mathrm{H}_{\mathrm{tot}}},
	\end{eqnarray}
with the efficiency 

\begin{equation}
	\epsilon = \frac{4.87\times 10^{-2}}{1 + 4\times 10^{-3} \psi^{0.73}}+ \frac{3.65\times 10^{-2}\left(T/10^4\right)^{0.7}}{1+2\times 10^{-4}\psi}.
\end{equation}
		
\subsubsection{Chemical heating}		
Chemical heating from exothermic reactions that form H$_2$ both in the gas phase and on dust is included following \citet{Grassi2014}, \citet{Hollenbach1979}, and \citet{Omukai2000}.

\subsection{H$_2$ formation on dust}\label{sec:h2formation}
The H$_2$ formation on dust is particularly relevant, as it determines  the atomic-to-molecular transition in the gas phase, being one of the most efficient formation paths for H$_2$. In previous work \citep{Gnedin2009,Gnedin2011,Christensen2012} a constant rate originally obtained by \citet{Jura1975} was employed, with the addition of a clumping factor ($C_{\rho}$) aimed to address the subgrid nature of H$_2$ formation in unresolved  high-density regions, namely

\begin{equation}\label{eq:H2dustJura}
	R_f (\mathrm{H_2}) = 3\times 10^{-17} n_{\mathrm{H}_{\mathrm{tot}}} n_{\mathrm{tot}} Z/Z_\odot C_\rho~~\mathrm{cm^{-3} s^{-1}}.
\end{equation}
The constant $3\times 10^{-17}$  has been motivated by observational measurements. We note that other versions of this rate have been proposed which include an explicit temperature dependency via a sticking coefficient \citep{Tomassetti2015}.


\krome allows the user to employ different types of dust with a given Mathis, Rumpl, \& Nordsieck (MRN)--like distribution $\sim a^{-\alpha}$ (\citealt{Mathis1977}) and to customize the range of sizes that can also extend to the regime of the polycyclic aromatic hydrocarbons (PAHs). We note here that there are different ways to treat the PAHs \citep{LiDraine2001b,TielensBook}, and that the PAH distribution is usually different from a standard power-law \citep{Weingartner2001}. 

\krome employs the H$_2$ formation on dust based on \citet{Cazaux2009}. They provided the rates for carbon and silicon-based grains  as
\begin{equation}\label{eq:H2dust}
 R_f(\mathrm{H_2}) = \frac{\pi}{2} n_\mathrm{H} v_g \sum_{j\in[\mathrm{C,Si}]}\sum_i  n_{ij} a_{ij}^2 \epsilon_j(T, T_d^i)  S(T, T_d^i)\,,
\end{equation}
where each \ith bin of the two grain species (i.e. C and Si) contributes to the total amount of molecular hydrogen formation. In the above equation $n_\mathrm{H}$ is the number density of atomic hydrogen in the gas-phase, $v_g$ is the gas thermal velocity, $n_{ij}$ is the number density of the \jth dust type in the \ith bin, $a_{ij}$ is its size. 
$T$ and $T_d^i$ are the temperatures of the gas and of the dust in the \ith bin, respectively.
The function $\epsilon_j$ has two expressions depending on the type of grain considered and has been reported by \citet{Cazaux2009}.
The sticking coefficient is given as
\begin{equation}
        S =\left[{1+0.4\,\sqrt{T_2 + \frac{T_d}{100}}+0.2\,T_2 +0.08\left(T_2\right)^2}\right]^{-1},\\
\end{equation}
with $T_2 \equiv T/(100\,\mathrm{K})$, according to \citet{Hollenbach1979}. We note that a clumping factor can also be applied to this rate.

As we are not interested in following the dust evolution (i.e. sputtering, growth, and shuttering processes are neglected), 
we can tabulate the H$_2$ formation on dust improving the performances of the code without losing in accuracy.


The final rates  are provided as tables using the following expression
\begin{equation}
R_f ({\rm H_2}) = \mu\,n_\mH\,f\mkern-6mu\left(T, n_{\mathrm{tot}}\right)n_{\mathrm{tot}},
\end{equation}
with $\mu$ being the mean molecular weight.
The details of the dust tabulation will be discussed in a forthcoming paper \citep{Grassi2016} where we will show a series of 3D hydrodynamic applications.

For this work we have prepared tables at a constant dust temperature $T_\mathrm{d} = 10$ K following the results discussed by \citet{Richings2014}, but a time-dependent $T_\mathrm{d}$ table can also be provided on request. We show in the Appendix that the on-the-fly approach, in which $T_\mathrm{d}$ is evaluated by solving the thermal balance equation (see \citealt{Grassi2014}), produces identical results to the case when $T_\mathrm{d}$ is kept constant. At the low densities considered in this work, $T_\mathrm{d}$ is not strongly affected by the interaction between dust and gas.

\subsection{Adiabatic index and mean molecular weight}
When coupling chemistry with hydrodynamics, different caveats arise. Two of the prevailing problems are the correct treatment of the adiabatic index ($\gamma$) and the mean molecular weight ($\mu$). Both quantities are strongly dependent on the chemical composition of the gas and are used to convert for instance pressure $\rightarrow$ energy $\rightarrow$ temperature through the ideal gas equation of state,

\begin{eqnarray}
	p &=& (\gamma -1) e,\\
	e &=& \frac{k_B T}{(\gamma -1)\mu m_{\mathrm{H}}},
\end{eqnarray}
 where $m_{\mathrm{H}}$ is the hydrogen mass, and $e$ the specific energy per unit mass. 
 
 \krome is able to evaluate the adiabatic index in different ways, allowing a large flexibility to the users. A simple temperature-independent approach has been presented in \citet{Grassi2014} and a more detailed method employs the calculation of the roto-vibrational partition functions, which provides a temperature-dependent adiabatic index (see \url{https://bitbucket.org/tgrassi/krome/wiki}). As we work in a low-density regime and we want to keep the approach simple, here we have decided to employ a constant adiabatic index, namely $\gamma = 5/3$, typical of a monoatomic or a collisionless gas. 
 
We use the $\mu$ computed by \krome based on the species abundances to have a fully consistent model.
 Our final mean molecular weight  is the summation of different contributions:

\begin{eqnarray}
	\mu &=& \frac{1}{\mu_e} + \frac{1}{\mu_{\mathrm{H}}} + \frac{1}{\mu_{\mathrm{He}}}  + \frac{1}{\mu_{\mathrm{metals}}},\\
	\mu_e &=&  \frac{x_e m_{\mathrm{H}}}{m_e},\\
   \mu_{\mathrm{H}} &=& x_{\mathrm{H}} + x_{\mathrm{H^-}} + 
   x_{\mathrm{H^+}} \frac{1}{2} (x_{\mathrm{H_2}} + x_{\mathrm{H_2^+}}), \\
   \mu_{\mathrm{He}} &=& \frac{1}{4}(x_{\mathrm{He}} + x_{\mathrm{He^+}} + x_{\mathrm{{He^{++}}}}),\\
   \mu_{\mathrm{metals}} &=& \frac{Z}{17.6003},
\end{eqnarray} 
 where $x_j$ are the species mass fractions, $Z$ is the metallicity in mass fraction, and we have assumed a mean mass number (number of protons+neutrons) for metals $\langle A\rangle_{\mathrm{metals}} = 17.6003$. If other He-based or H-based species are added to the chemical network the above definitions of the individual mean molecular weight should be updated. This is usually done automatically by \krome during the pre-processing stage.
 
\section{Testing the models}\label{sec:models_test}
In the following we show how the different ingredients of our model affect the chemical and thermal evolution of the gas. We distinguish between two classes of models (see Table \ref{tab:models}): (i)  Model I and Model II, without any radiation background, and (ii) Model III and Model IV, which include the radiation background of \citet{Haardt2012} and can be employed to study the chemical evolution in galaxy simulations and follow the HII-HI-H$_2$ phase transitions. 
In the following tests we will use a simple one-zone cloud collapse model \citep*{Omukai2000}, which evolves the density $\rho$ based on the free-fall time $t_{\mathrm{ff}}$,

\begin{equation}\label{eq:one-zone}
	\frac{d\rho}{dt} = \frac{\rho}{t_{\mathrm{ff}}},\\
\end{equation}
\noindent with $t_{\mathrm{ff}} = \sqrt{3\pi/32 G \rho}$, where $G$ is the gravitational constant. The thermal evolution is solved together with Eq. (\ref{eq:one-zone}) as

\begin{equation}\label{eq:dT}
	\frac{dT}{dt} = \frac{(\Gamma - \Lambda) (\gamma - 1)}{k_B n_{tot}}.
\end{equation}
We have already shown \citep{Grassi2014} that \krome is able to reproduce previous results obtained by employing the above one-zone framework \citep{Omukai2000}, hence we consider this problem a reliable test to explore the physics discussed in this work.
The choice of performing one-zone cloud collapse tests is also due to the possibility to dynamically explore the chemical/thermal evolution at different densities, to resemble the transition through the different gas phases. In Section \ref{sec:metal_noneqCool} we additionally present tests at fixed densities, i.e. a time evolution of an isochoric gas.

\subsection{The neutral models}\label{sec:model1}
Model I and Model II  in Table \ref{tab:models} are based on low-complexity networks, and include approximate metal cooling, i.e. a network composed of a mixture of H/He-based species, with the equilibrium metal tables discussed in Section \ref{sec:thermal}. The two models employ different H$_2$ formation rates on dust: (i) Model I employs the approximated H$_2$ formation rate on dust (Eq. \ref{eq:H2dustJura}),  which does not depend on the dust composition or the dust size distribution and is widely used in galaxy simulations, and (ii) Model II  employs an improved dust treatment including dust tables evaluated as discussed in Section \ref{sec:h2formation}. These tables assume 40 bins in size of dust grains of mixed composition including carbonaceous (20 bins) and silicates (20 bins) with a fixed dust temperature \mbox{$T_\mathrm{d}$ = 10 K}. The metallicity is kept constant as $Z = 0.5~ Z_\odot$, and we will show how the metallicity affects the results in Section~\ref{sec:metalparam}.
We evolve the one-zone collapse starting with the following initial conditions: $T = 100$ K, $n_{\mathrm{tot}}$ = 1 cm$^{-3}$, $n_{\mathrm{H}} = 0.92~n_{\mathrm{tot}}$, $n_{\mathrm{He}} = 0.0755~n_{\mathrm{tot}}$, $n_{\mathrm{H^+}} = n_{\mathrm{e}} = 10^{-4} n_{\mathrm{tot}}$, $n_{\mathrm{H_2}} = 10^{-6} n_{\mathrm{tot}}$, which we evolve until a density of 10$^6$ cm$^{-3}$ is reached. The other species abundances are set to zero.
For Model II we use a standard MRN dust distribution ($a^{-3.5}$) which scales as $D = D_\odot\times Z/Z_\odot$, where $D_\odot = 0.00934$ is the dust-to-gas mass ratio at solar metallicity. In addition to this we evolve the same Model II, but including dust cooling (referred to as Model IIa) to assess the impact of this contribution on the thermal evolution.

\begin{figure}
	\begin{center}
	\includegraphics[scale=0.7]{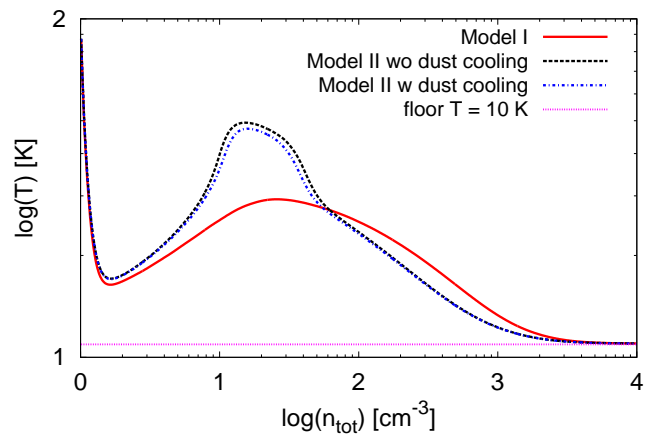}
	\caption{Thermal evolution versus the total number density for the three models described in Section \ref{sec:model1}. The red curve refers to Model I, which employs a simple prescription for the H$_2$ formation rate on dust. The black curve employs the more accurate \citet{Cazaux2009} rate for H$_2$ formation on dust, and the blue curve is the same Model II but including also the dust cooling as reported in Eq~\ref{eq:dustcool}.  The magenta line represents the CMB floor at $z = 3$.}\label{fig:model1}
	\end{center}
\end{figure}

\begin{figure}
	\begin{center}
	\includegraphics[scale=0.7]{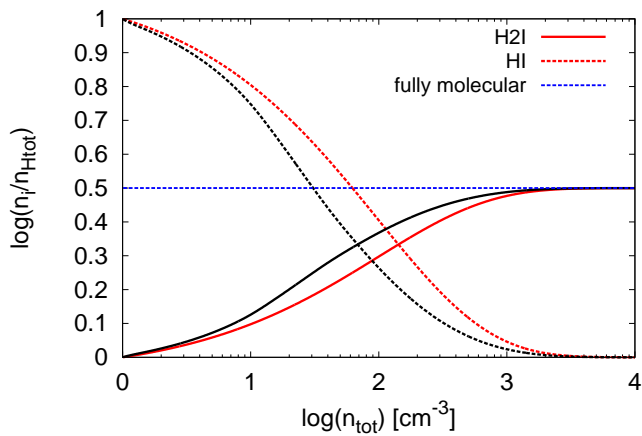}
	\caption{H$_2$ and H number fraction as a function of the total number density. Red curves refer to Model I, while black curves to Model II. Note that Model II with dust cooling is not reported here as it overlaps with the case without dust cooling.}\label{fig:model2}
	\end{center}
\end{figure}

\begin{figure}
	\begin{center}
	\includegraphics[scale=0.7]{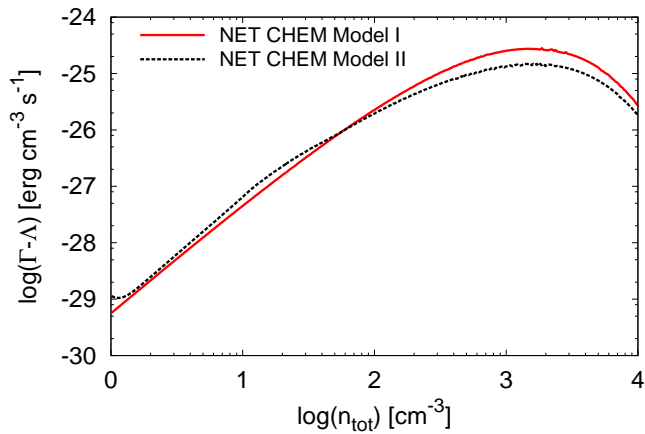}
	\caption{H$_2$ net chemical heating as a function of the total number density. The red curve represents Model I, while the black curve refers to Model II.}\label{fig:model3}
	\end{center}
\end{figure}

In Fig. \ref{fig:model1} we report the thermal evolution of the gas as a function of the total number density for the three models presented above. The H$_2$ formation on dust computed with a proper dust model produces a  hotter gas around 10$^1$~-~10$^2$~cm$^{-3}$. This is mainly due to the fact that more heating is produced from the formation of H$_2$ caused by an increase in the formation efficiency due to the rate by \citet{Cazaux2009}. The increase is also clear from Fig. \ref{fig:model2}, where the evolution of atomic and molecular hydrogen (in terms of $n_i/n_{\mathrm{H}_{\mathrm{tot}}}$) is shown for the two models. In Fig. \ref{fig:model3} we report the net H$_2$ formation heating  $|\Gamma - \Lambda|$. In the case where a proper dust model is included the transition HI-H$_2$ occurs a bit earlier.
The effect of dust cooling (Model IIa) at these densities is in practice negligible (Fig. \ref{fig:model1}).

\subsection{Models irradiated by UVB/X-ray photons}
\begin{figure}
	\begin{center}
	\includegraphics[scale=0.68]{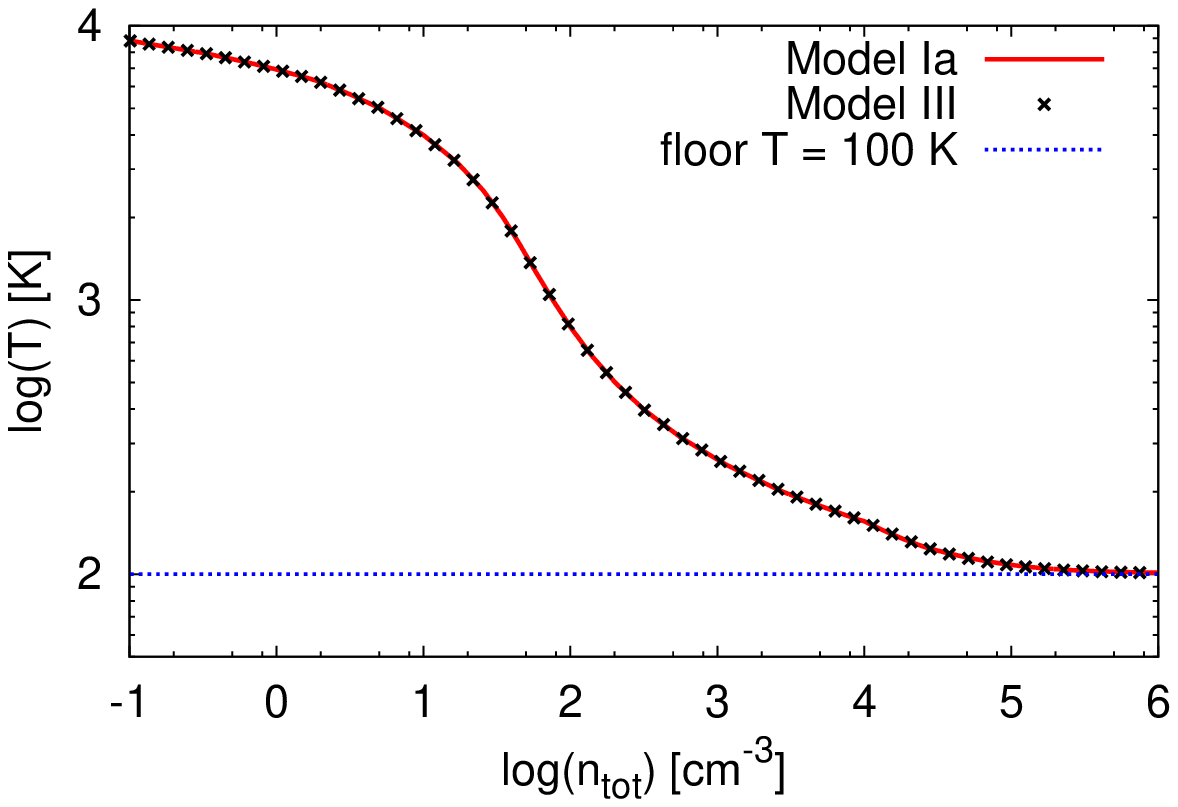}\\
	\includegraphics[scale=0.72]{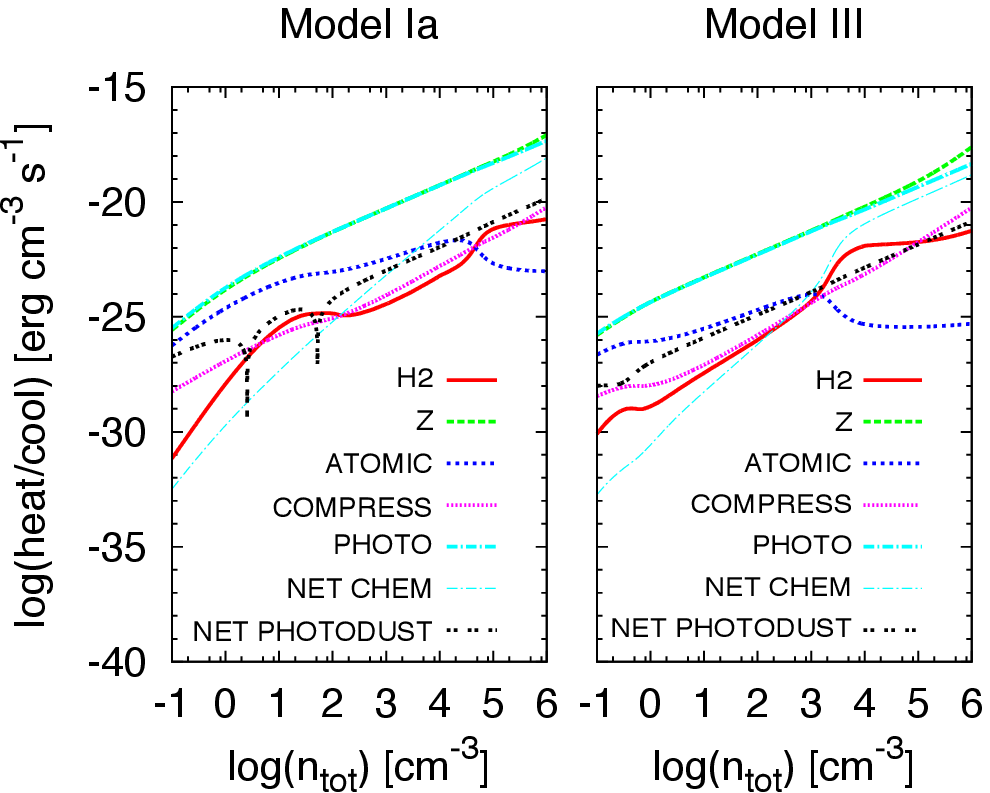}
	\caption{Top: thermal evolution as a function of the total number density for Model Ia (with a clumping factor $C_{\rho}$ = 1) and Model III. Bottom: heating and cooling functions versus the total number density for Model Ia (left) and Model III (right). We omit here the model with clumping factor $C_{\rho}$ =10, as this is identical to Model Ia.}\label{fig:model4}
	\end{center}
\end{figure}

We now explore what happens to a gas of metallicity $Z = 0.5~~Z_\odot$ under a UVB/X-ray radiation background computed at redshift $z = 3$. In the first of these models (Model III in Table \ref{tab:models}) we include the H/He photo-processes reported in Table \ref{tab:photo} and photo-heating, together with a proper treatment of the photoelectric heating by dust. This model is compared to Model I discussed in the previous Section, which includes the approximate H$_2$ formation rate on dust. Here we have added the photochemistry for consistency and refer to it as Model Ia. An additional parameter in Model Ia is the clumping factor $C_{\rho}$, for which we explore the results of $C_\rho$~=~1 and $C_\rho$ = 10, the latter enhancing the H$_2$ formation on dust by an order of magnitude. We assume here a gas at $T = 2\times 10^4$ K, with a slightly lower density compared to Models I and II (i.e. $n_{\mathrm{tot}} = 0.1$ cm$^{-3}$), and a fully ionised chemical composition. The results are reported in Fig. \ref{fig:model4}. 

Two important pieces of information can be retrieved from this test: (i) under the conditions considered in this simple test, having an approximate  H$_2$ formation rate on dust or a more accurate treatment does not affect the thermal evolution; (ii) changing the clumping factor in Model Ia, i.e. increasing the H$_2$ formation rate on dust by an order of magnitude has also a negligible effect on the final thermal evolution. For this reason we only report in Fig. \ref{fig:model4} a comparison between Model Ia ($C_{\rho}$ = 1) and Model III. This behaviour can be better understood if we look at the different heating/cooling contributions for the evolution of the gas irradiated by a UVB/X-ray background (bottom panels of Fig.~ \ref{fig:model4}). The thermal evolution in both models is dominated by photo-heating and metal line cooling, which balance each other, while the net chemical heating has no influence at all on the final evolution. This is the effect of the ionising background which also produces a warmer gas. Hence, the effect of having a different H$_2$ formation rate on dust is negligible for the thermal evolution. The atomic cooling controls the evolution together with the photoelectric heating by dust. 
 
\begin{figure}
	\begin{center}
	\includegraphics[scale=0.7]{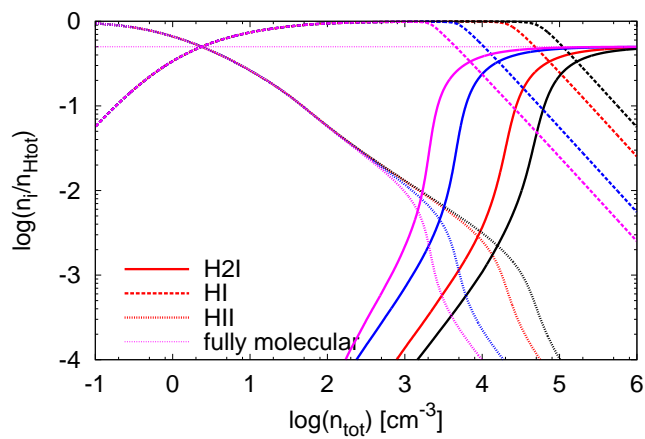}
	\caption{H$_2$ and H number fraction as a function of the total number density. Black curves refer to Model Ia,  red curves to Model III,  blue curves to Model Ia but with a clumping factor  $C_{\rho}$~=~10, and magenta curves to Model III but with the H$_2$ formation rate 10 times larger.}\label{fig:model5}
	\end{center}
\end{figure}

 However, when we look at the HII-HI-H$_2$ transition (Fig. \ref{fig:model5}), to change the H$_2$ formation rate on dust produces a difference. For instance, employing an accurate and appropriate grain size-dependent approach (Model III) leads to the transition at lower densities compared to the approximated rate (black curves in the plot). The gas becomes fully molecular ($n_{\mathrm{H_2}}/n_{\mathrm{H}_{\mathrm{tot}}}$ = 0.5)  earlier. If we  increase the clumping factor by an order of magnitude, then also the density at which the gas becomes fully molecular is reduced by an order of magnitude. This holds both for the approximate rate (Model Ia) and the rate provided by \citet{Cazaux2009}. We conclude that it is very important to include a proper dust treatment (grain size-dependent) to accurately assess the atomic-to-molecular transition in galaxy evolution simulations. We also note that the clumping factor is making a large difference. It is quite common in hydrodynamic simulations of galaxy evolution to set this parameter to $C_{\rho}$=10 or even higher values     
  \citep{Gnedin2009}, inducing a faster HI-H$_2$ transition to account for unresolved high-density gas. If the need for such a clumping factor however prevails also in higher-resolution simulations, it may also point towards an underestimate of the H$_2$ formation rates. 
 At the same time, we note that the HII-HI transition is completely dominated by photo-chemistry and results in an identical evolution for the three models discussed in this Section.
  
\begin{figure}
	\begin{center}
	\includegraphics[scale=0.7]{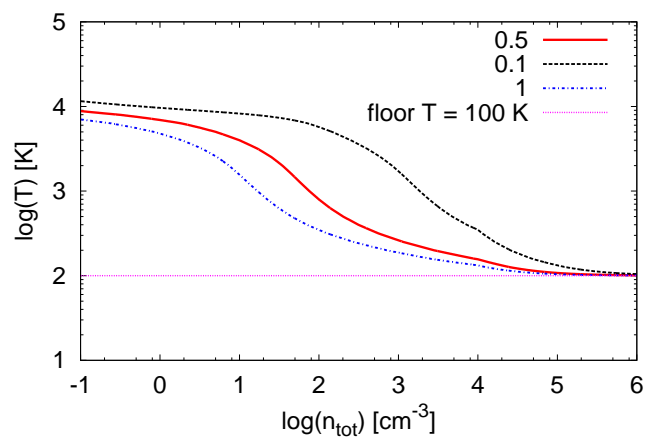}
	\caption{Thermal evolution for Model III at three different metallicities, namely $Z = 0.1~Z_\odot$,  $0.5~Z_\odot$, and $Z_\odot$ (solar metallicity). The temperature floor is sketched by the magenta line.}\label{fig:metal1}
	\end{center}
\end{figure}
\subsubsection{Effects of the radiation background and metallicity}\label{sec:metalparam}
As already reported by \citet{Richings2014}, the type of radiation background, its strength and the metallicity can affect the evolution of the gas (and, subsequently, the star formation) as they change the ionisation processes and the dust/metal related physics.

\begin{figure}
	\begin{center}
	\includegraphics[scale=0.7]{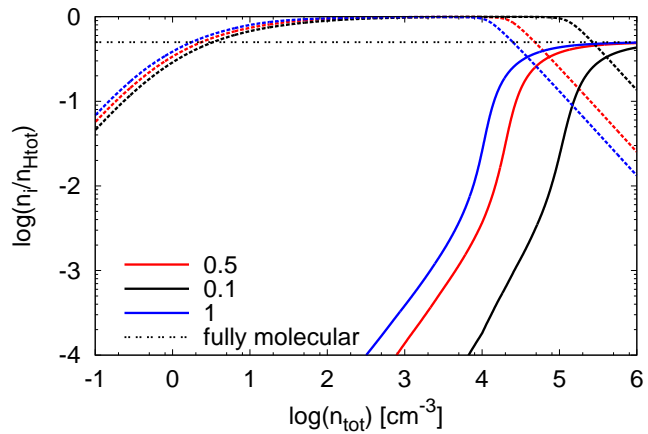}
	\caption{Atomic and molecular hydrogen evolution for three different metallicities, $Z = 0.1~Z_\odot$, $0.5~Z_\odot$, and $Z_\odot$ (solar metallicity), for Model III. The fully molecular stage is marked by the horizontal dotted line.}\label{fig:metal2}
	\end{center}
\end{figure}

Here we present three models with different metallicities, namely $Z/Z_\odot$ = 0.1, 0.5, and 1 (i.e. solar metallicity). The metallicity both affects the metal cooling and the H$_2$ formation on dust, which have a linear dependence on it. In Fig.~\ref{fig:metal1} we see the boosting of metal cooling when going from  lower to  higher metallicities. The effect of metallicity on H$_2$ formation reported in Fig. \ref{fig:metal2} plays a relevant role, being more evident when we switch from 0.1 $Z_\odot$ to solar metallicity with a difference of two orders of magnitude in the density level at which the fully molecular stage is reached. Changing the metallicity has then the same effect as enhancing the rate of H$_2$ formation on dust.

\begin{figure}
	\begin{center}
	\includegraphics[scale=0.66]{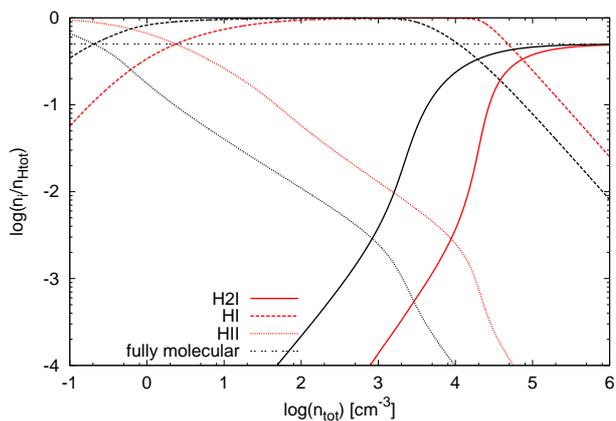}
	\caption{Ionised, atomic, and molecular hydrogen evolution as a function of the density for two different radiation backgrounds, the standard extragalactic background at redshift $z=3$ (red curves) by \citet{Haardt2012}, and 10 per cent of the latter (black curves). The results come from the evolution of Model III.}\label{fig:metal3}
	\end{center}
\end{figure}

\begin{figure}
	\begin{center}
	\includegraphics[scale=0.66]{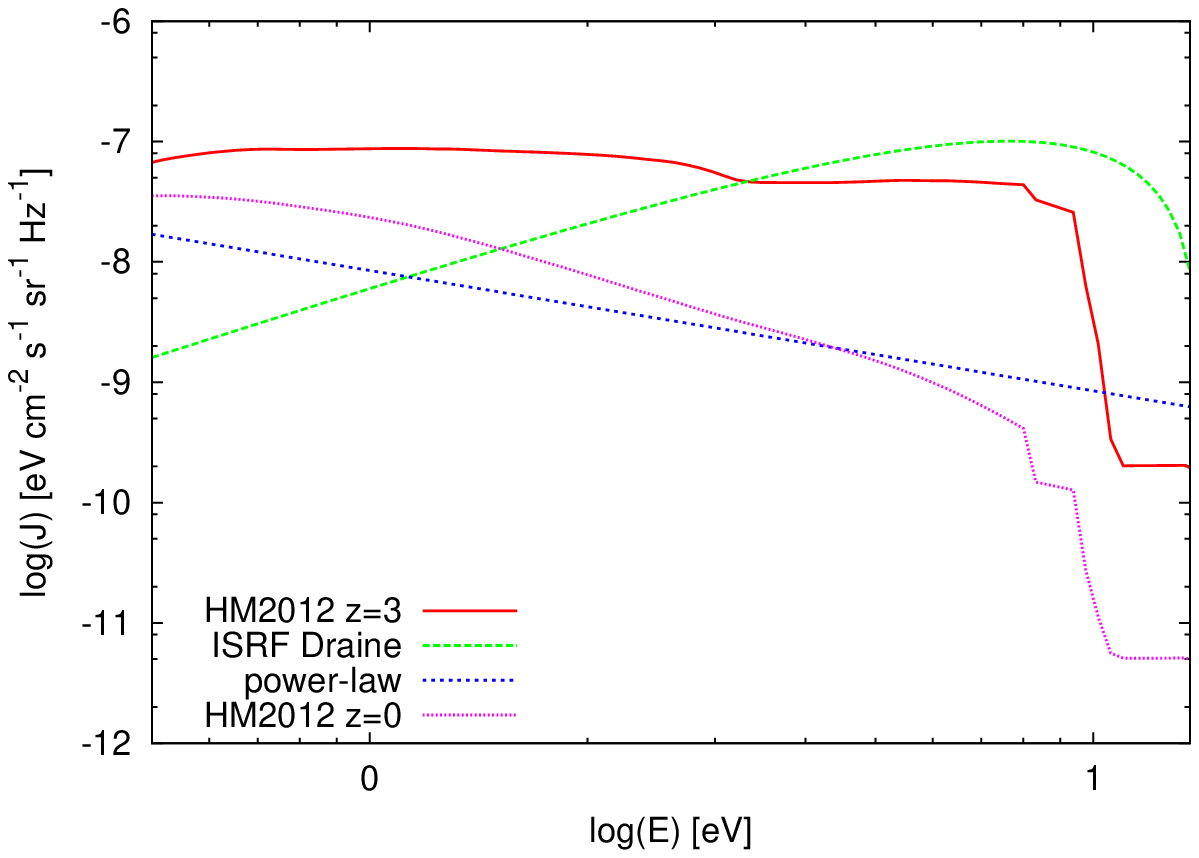}
	\caption{Different radiation backgrounds as a function of energy up to 13.6 eV. Red solid curve and magenta curve: extragalactic radiation background from \citet{Haardt2012} at redshift $z=3$ and $z=0$, respectively; green dotted curve: the standard ISRF by \citet{Draine1978}; blue dashed curve: a simple power-law spectrum. See text for details.}\label{fig:metal4}
	\end{center}
\end{figure}

As we are employing a very strong photo-ionising radiation background, it is important to assess the differences produced once we change this radiation background. Taking ten per cent of the extragalactic radiation background produces a significant effect in the  HII-HI-H$_2$ transitions, as fewer ionisations occur and the atomic phase is reached much earlier boosting the formation of H$_2$ at lower densities (Fig. \ref{fig:metal3}). This can be considered as mimicking the effect of  shielding produced by a radiative transfer algorithm which indeed will tend to strongly reduce the ionisation processes in dense regions.

We note here that the radiation background at redshift $z=3$ is quite strong and  comparable to the galactic radiation produced by stellar sources. We report in Fig. \ref{fig:metal4} a comparison of different radiation backgrounds for energies up to 13.6 eV: the extragalactic radiation by \citet{Haardt2012} at $z=3$ and $z=0$ (commonly employed in galaxy simulations), the interstellar radiation field (ISRF) by \citet{Draine1978}, and a typical power-law radiation background $(E/E_0)^{-1}$, with $E_0 = 13.6$~eV. The ISRF by \citet{Draine1978}  is comparable to the extragalactic radiation background by \citet{Haardt2012}, which is even stronger at lower energies.

 \subsection{The effect of non-equilibrium metal cooling}\label{sec:metal_noneqCool}
 \begin{figure}
	\begin{center}
	\includegraphics[scale=0.7]{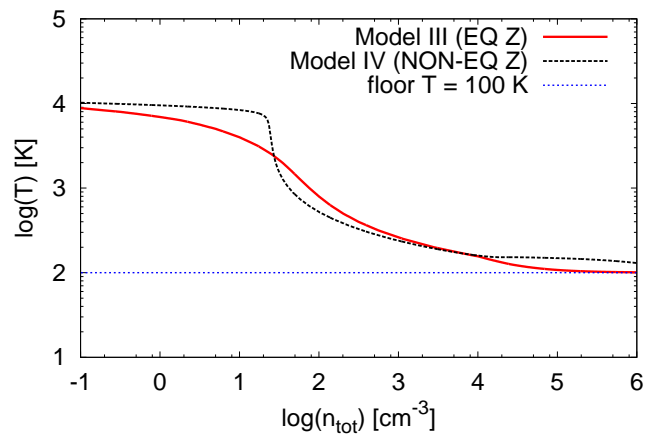}
	\includegraphics[scale=0.7]{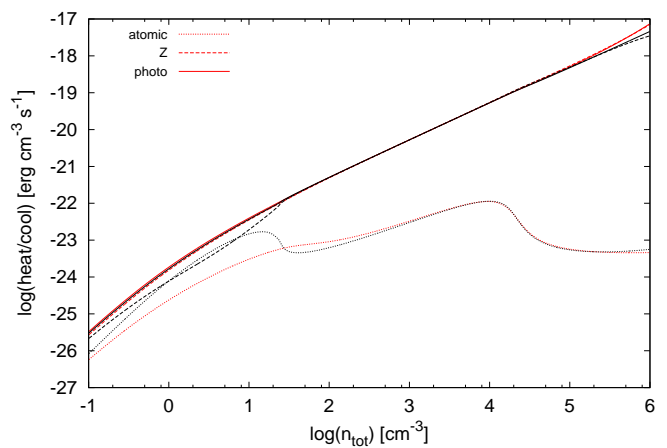}
	\caption{Top: thermal evolution for Model III (red) and Model IV (black) as a function of the total number density. Bottom: main cooling/heating functions for the two models. Red curves refer to Model III and black curves to Model IV. The different contributions reported in the figure are the following: atomic cooling (dotted lines), metal cooling (dashed lines), and photo-heating (solid lines).}\label{fig:model6}
	\end{center}
\end{figure}

\begin{figure}
	\begin{center}
	\includegraphics[scale=0.66]{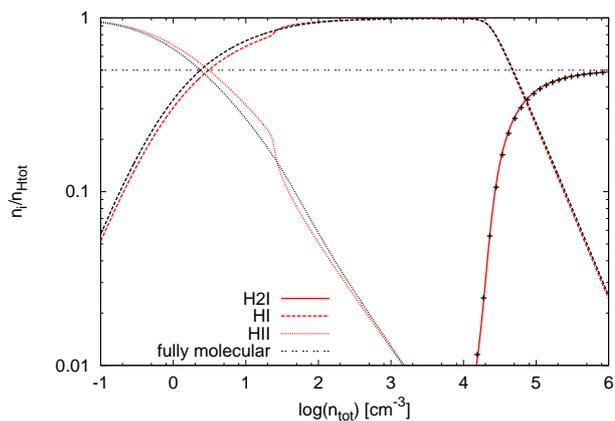}
	\caption{Chemical evolution of H$_2$, H$^+$, and H, as a function of the total number density for Model III, which employs the equilibrium metal cooling at all temperatures (red curves), and Model IV, which employs the non-equlibrium metal cooling for temperature  $T < 10^4$ K (black curves and crosses). The effect of non-equilibrium cooling on the H$_2$ evolution is negligible and produces the same results as the equilibrium cooling model (crosses). A horizontal line at $n_i/n_{\mathrm{H_{tot}}}$ = 0.5 is sketched to represent the fully molecular stage.}\label{fig:model7}
	\end{center}
\end{figure}

\begin{figure}
	\begin{center}
	\includegraphics[scale=0.66]{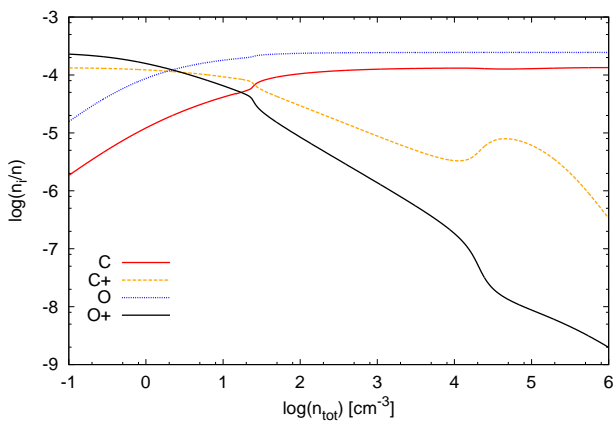}
	\caption{Most relevant metal species number fraction evolution (C, C$^+$, O, and O$^+$) as a function of the total number density.}\label{fig:model8}
	\end{center}
\end{figure}
 As already discussed in Section \ref{sec:thermal}, non-equilibrium metal line cooling, i.e. solving at run time the linear system for the excitation levels of every single metal, can have an impact on the evolution of the gas at temperatures $T < 10^4$~K. In Model IV, we evolve the H/He mixture network plus  seven additional metal species (C, C$^+$, O, O$^+$, Si, Si$^+$, and Si$^{++}$) and reactions involving these metals as listed in Table \ref{tab:metals}. 
 We have in the end a network with 16 species and 74 reactions that we evolve with conditions similar to the one employed in the previous tests, i.e. a fully ionised gas at $T = 2\times 10^4$ K. Note that the metal number densities are initialised rescaling the solar abundances by the metallicity $Z/Z_\odot$. The main difference between Model III (discussed in the previous Section) and the current Model IV is  the treatment of the metal cooling below 10$^4$ K, where  we  employ Eq. (\ref{eq:metaltot}) for the non-equilibrium treatment.
 
 In the top panel of Fig. \ref{fig:model6} we show the comparison between Model III and Model IV, to see how much the thermal evolution is affected by the different treatment of metal line cooling. When we employ the non-equilibrium approach we have less net cooling. If we look at the bottom panel of Fig. \ref{fig:model6} we see a difference in the two metal cooling functions of about 10 per cent, which increases at higher densities ($n_{\mathrm{tot}} = 1 - 10^2$ cm$^{-3}$), then the two contributions equate each other as the collisions dominate over the photo-ionisations, as already discussed in Section \ref{sec:thermal}.
 
In Figs. \ref{fig:model7} and \ref{fig:model8} we report the evolution of the hydrogen  and metal species as a function of density, respectively.
When we employ the  non-equilibrium cooling, the evolution of the hydrogen species (and in particular the HI-H$_2$ transition) is not affected at all, as the cooling is only changing the thermal evolution at low densities. We argue that there is only a minor effect mainly in the HII-HI transition. However, it could be important to assess how much the dynamics and the structure of the galaxy are affected in a realistic 3D simulation. Fig. \ref{fig:model8} shows how quickly the most relevant metal ions recombine. This occurs at high densities when the temperature is lowered and recombinations become faster.

 

\begin{figure}
	\begin{center}
	\includegraphics[scale=0.7]{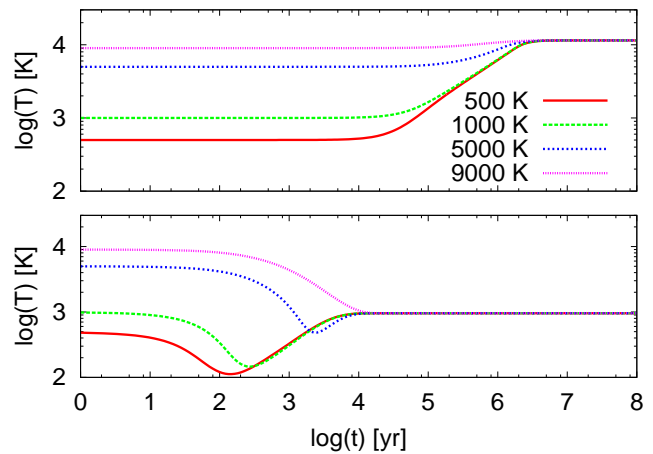}
	\caption{Gas thermal evolution for two different densities, 0.1~cm$^{-3}$ (top panel), and 100 cm$^{-3}$ (bottom panel), and starting from four different temperatures, namely 500, 1000, 5000, and 9000 K. The system is evolved until it reaches the thermo-chemical equilibrium.}\label{fig:comparisonZ2}
	\end{center}
\end{figure}

 
For a self-consistent comparison between a non-equilibrium and a photo-ionisation equilibrium approach we prepared a test which evolves (below 10$^4$ K) the full system of ODEs for a long enough time to reach the thermo-chemical equilibrium by employing our non-equilibrium metal cooling, i.e. the system is evaluated on the fly. In this way we can provide important information about the time scale to reach equilibrium and understand when an equilibrium approach is correct and when it is leading to an overestimate or underestimate of the cooling. For reference we take a typical integration time-step for galaxy simulations which can be estimated to be $\sim$ 10$^4$ yr, even if we should consider that a large change in the thermochemical conditions may reduce the hydrodynamical time-step in order to take into account their effect.
We evolve the system for two different constant densities (isochoric evolution) and four different initial temperatures and show the results in Fig.~\ref{fig:comparisonZ2}. As this is a thermochemical equilibrium, the system will reach the same final equilibrium temperature $T_{eq}$ for the same gas density.

The equilibrium is reached at different times and this obviously depends on the density: a higher density gas reaches the equilibrium much earlier. 
If we consider a typical numerical time-step of 10$^4$ yr, it is clear that assuming equilibrium for low-density gas leads to a large error in the cooling as the equilibrium needs at least 1 Myr to be reached. However, in situations where the density is high, the time for the system to reach  equilibrium is comparable to 10$^4$ yr and then this assumption becomes valid on small scales.

\begin{figure}
	\begin{center}
	\includegraphics[scale=0.7]{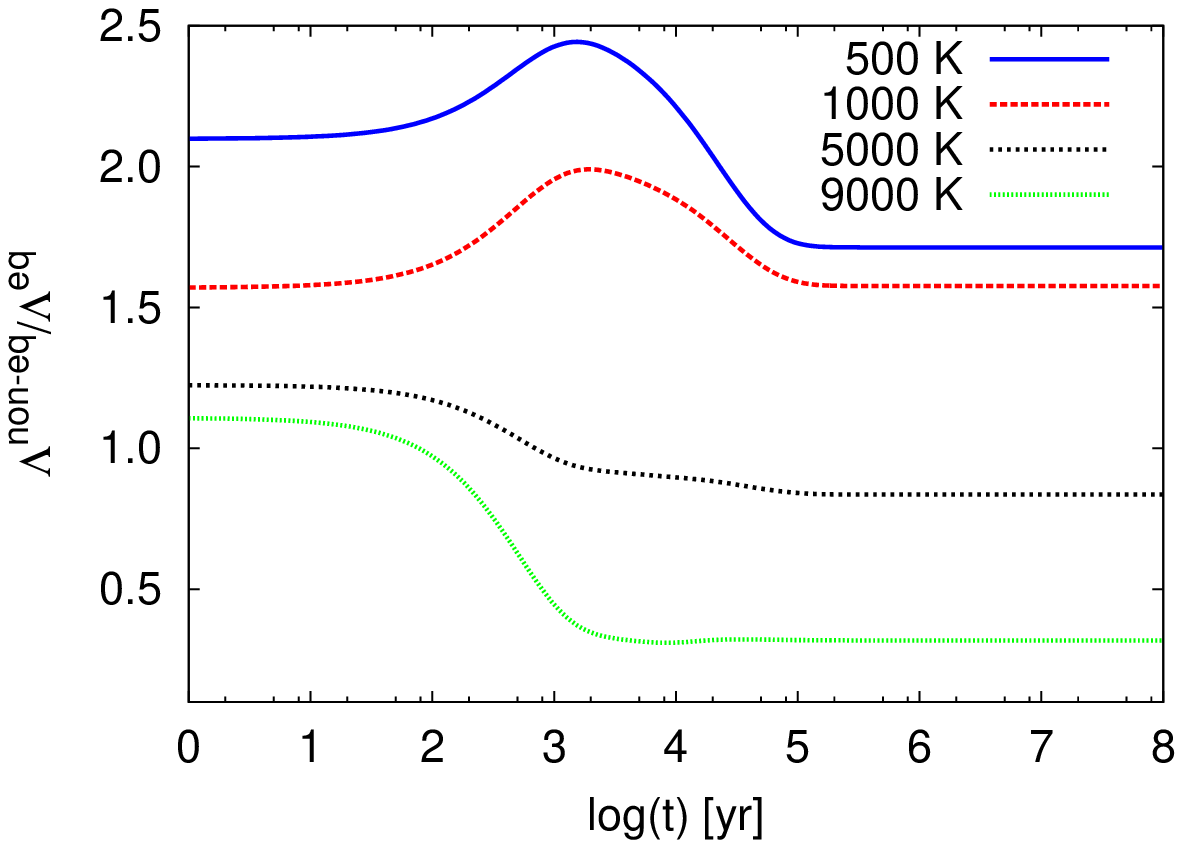}\\
	\includegraphics[scale=0.7]{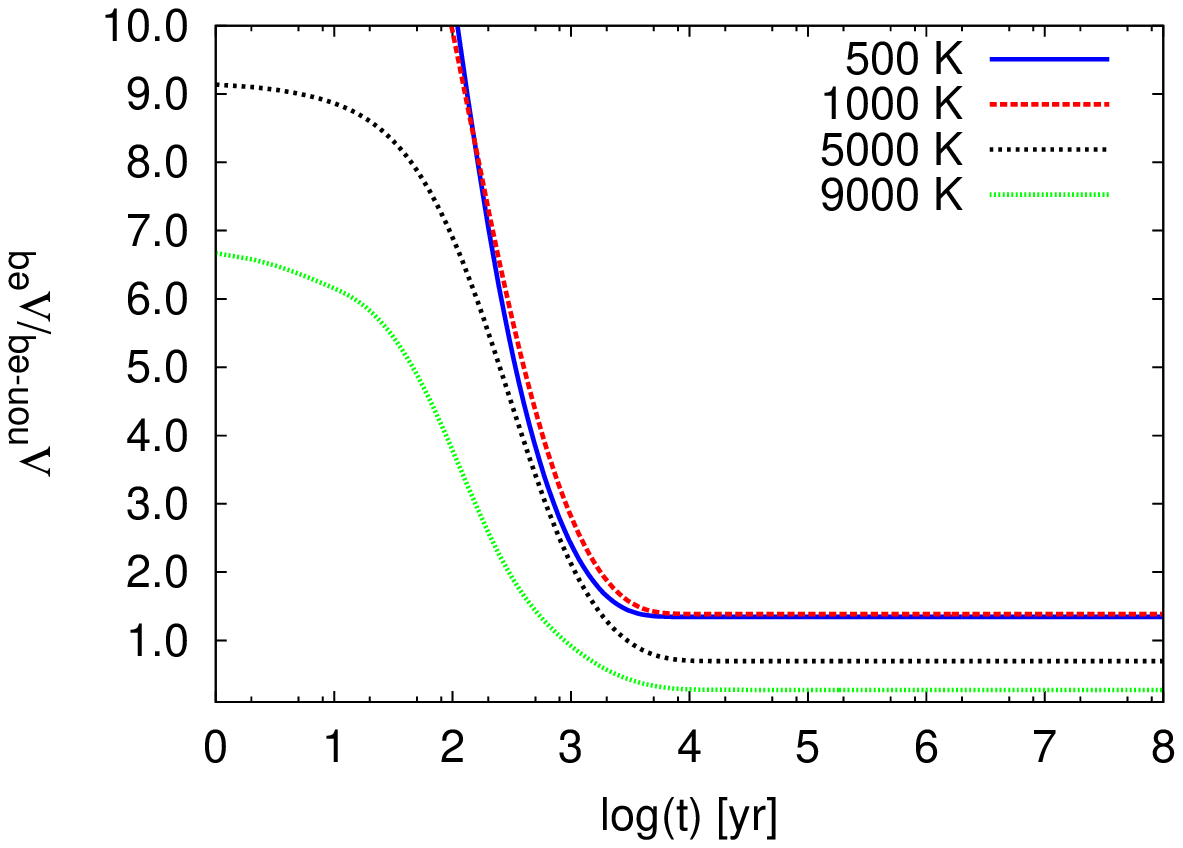}
	\caption{Time evolution of the non-equilibrium/equilibrium metal cooling ratio for two different densities, namely 0.1 cm$^{-3}$ (top panel), and 100 cm$^{-3}$ (bottom panel), for four different temperatures as reported in the legend.}\label{fig:comparison_eqnon}
	\end{center}
\end{figure}
As a second test, we evolve the system with a \textsc{cloudy}-like approach, i.e. not only keeping the density constant but also the temperature and evolving the system until it reaches the chemical equilibrium. The aim of this test is to quantify the differences between the non-equilibrium and the equilibrium cooling, and in particular how far our approach is from the standard \textsc{cloudy} equilibrium tables (PIE). In Fig. \ref{fig:comparison_eqnon} we show the ratio $\Lambda^{\mathrm{non-eq}}/\Lambda^{\mathrm{eq}}$ for two different constant densities, 0.1 cm$^{-3}$ (top panel), and 100 cm$^{-3}$ (bottom panel), and four different initial temperatures, 500, 1000, 5000, and 9000 K, respectively. 
It is worth noting that  there are cases where assuming the equilibrium might lead to a large overestimate/underestimate of the final metal cooling.
From this simple test we can infer that: (i) the non-equilibrium cooling for temperatures below 5000 K is larger than the equilibrium cooling by a factor of two and for high densities by orders of magnitude. This can be due  to the fact that our approach includes collisions with neutral species like H and H$_2$, while the \textsc{cloudy} tables assume the collisions to be dominated by electrons; (ii) when we compare the equilibrium reached with our approach versus  \textsc{cloudy}, i.e. at the plateau (for times larger than 1 Myr or less depending on the density and temperature), we see that the differences are about a factor of two. This behaviour may be due to the different number of metals, transitions, and colliders between our approach and the \textsc{cloudy} one. Note  also that the chemical network is different. In most cases, the time to reach  equilibrium is larger than a typical hydrodynamical time-step for temperatures below 1000 K and a non-equilibrium approach is then desirable. Our results are also in agreement with recent similar tests reported by \citet{Richings2014}, who have shown that non-equilibrium cooling is usually enhanced compared to the equilibrium case. It is important to note that we are not comparing the net cooling here, as we are interested on the differences between the metal cooling only. Obviously the photo-heating in the non-equilibrium and equilibrium cases will be different, as well as other cooling/heating contributions that can change the final real effect on the thermal evolution of the gas, as for instance in the test reported in Section~\ref{sec:metal_noneqCool}, which provided less net cooling in the non-equilibrium case. 


 \subsection{Radiation attenuation and caveats}\label{sec:shield_caveats}
 In the tests presented in the previous sections, we focussed on the optically thin case employing only the extragalactic radiation background by \citet{Haardt2012}. For a realistic study of galaxy properties it is necessary to include a proper radiative transfer module which follows the history of photons during the simulation. A local UV radiation background produced by the stars is also a desirable ingredient and can be much stronger than the extragalactic background (see discussion in Section \ref{sec:metalparam}). This piece of information should therefore come from a more accurate radiative transfer treatment. The computational cost of coupling radiative transfer with chemistry is not negligible and over the years  local approximations have been proposed. 
 
 As this problem does not concern the model presented here and pursued with the package \kromes, we will now only perform some simple tests to illustrate some of the results. For this purpose, we will assume that the UV flux provided to \krome has already been obtained from a radiative transfer calculation, so that the UV radiation can be treated in the optically thin approximation. For the photodissociation of H$_2$, self-shielding and shielding by dust may however be relevant, and we explore here how much it matters within the one-zone framework.
 
An optical depth is commonly applied to the optically thin rates assuming $\tau = \sum_i \sigma_i N_i$, with $\sigma_i$ being the photo cross-section of the \ith species and $N_i$  the column density, which is evaluated in different ways and in general depends on a characteristic length, 
\begin{equation}
	N_i = n_i L_i,
\end{equation} 
where $L_i$ is often based on a velocity or density gradient (i.e. a Sobolev-like length) or on the Jeans length, as in the tests reported in this work.

\begin{figure}
	\begin{center}
	\includegraphics[scale=0.7]{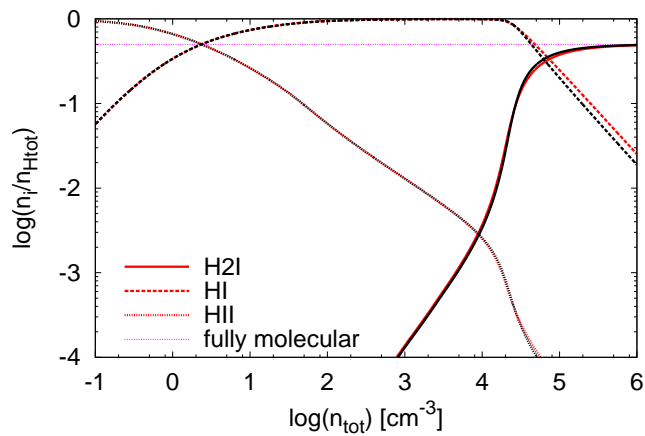}
	\caption{Chemical evolution of H$_2$, H$^+$, and H, as a function of the total number density $n_{\mathrm{tot}}$, for Model III with and without self-shielding as discussed in Section \ref{sec:shield_caveats}. A line at $n_i/n_{\mathrm{H_{tot}}}$ = 0.5 is sketched to represent the fully molecular stage.}\label{fig:shielding}
	\end{center}
\end{figure}

We focus on H$_2$ and dust shielding which provide an optically thick rate of
 
 \begin{equation}
 	k_{ph}(\mathrm{H_2})^{\mathrm{thick}} = k_{ph}(\mathrm{H_2})^{thin} S_d(\mathrm{H_2}) S_{\mathrm{self}}(\mathrm{H_2}),
 \end{equation}
 where $S_{\mathrm{self}}$ is adopted by \citet{Wolcott2011}, and $S_d$ is defined following \citet{Richings2014} as

\begin{equation}
	S_d(\mathrm{H_2}) = \exp(- \sigma_d \gamma_{H_2} N_{H_{tot}} Z/Z_\odot),\\
\end{equation}
with $\gamma_{H_2} = 3.74$, $N_{\mathrm{H_{tot}}} = N_{\mathrm{HI}} + N_{\mathrm{HII}} + 2 N_{\mathrm{H_2}}$,  and $\sigma_d = 4.0\times 10^{-22}$ cm$^2$. 
We note here that \citet{Gnedin2009} and \citet{Gnedin2011} use different values for $\sigma_d$. We assume here a column density based on the Jeans length $\lambda_J$, which is suitable for the one-zone collapse problem:

\begin{equation}
N_i = \frac{1}{2} n_i \lambda_J.
\end{equation}
As we show in Fig. \ref{fig:shielding}, the shielding is not relevant for our test-case. This is mainly due to the fact that in the presence of a fixed ionising background, photo-heating is very strong and the high temperature delays the formation of molecular hydrogen, as collisional dissociation is an efficient destruction channel above $2000$~K. 

It is important to note here that the above formulation of the self-shielding comes from a static slab of gas and that the Jeans length generally overestimates the column density \citep{Wolcott2011}. To properly evaluate the column density and then the H$_2$ shielding we would need to probe the relative velocity between particles and compare it with the thermal velocity. This can be assessed with algorithms like for example \textsc{TreeCol} \citep{Clark2012,Hartwig2015}. However, this is again an issue of modelling radiation transfer that does not concern the microphysics and the chemistry solved by \kromes. \citet{Seifried2015} recently reported 3D calculations of collapsing ISM filaments where the \textsc{TreeCol} algorithm has been successfully coupled with \krome and we refer interested readers to this paper.

Moreover, as pointed out by \citet{Richings2014}, to apply an optical depth to the optically thin atomic photo-rates within the chemical solver is computationally unfeasible and it is rather inaccurate, as it does not consider the geometry, the photon history, and the environment. It is then crucial to have a radiation attenuation coming from the solution of the radiative transfer equation. In addition, some colleagues pursue more simple approaches, like the one presented by \citet{Christensen2012} and \citet{Tomassetti2015}, which provide reasonable results for the average atomic-to-molecular transition in galaxy simulations, while of course local variations exist and are expected within real galaxies.  

\begin{figure}
	\begin{center}
	\includegraphics[scale=0.7]{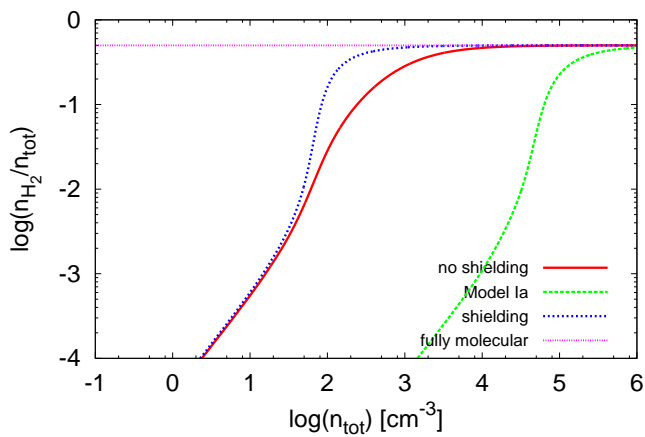}
	\caption{Chemical evolution for H$_2$ as a function of the total number density for different physical conditions, $C_\rho$ = 10, $Z$~=~$Z_\odot$ and $z=0$ without shielding (red curve), and with shielding (blue curve), and our reference model with $C_\rho$ = 1, $Z$~=~0.5~$Z_\odot$, $z=3$, labelled as Model Ia (green curve).}\label{fig:gnedin}
	\end{center}
\end{figure}

To provide a more reasonable test and to show the effect of the shielding on H$_2$, we perform  runs resembling the conditions often employed in galaxy simulations, i.e. solar metallicity, an extragalactic background corresponding to a redshift $z = 0$, and a clumping factor of at least $C_{\rho}$ = 10, all parameters that boost the formation of H$_2$ allowing its formation at densities where the shielding becomes important ($n_{\mathrm{tot}}\sim$ 10 cm$^{-3}$). In these tests we keep the approach simple and consistent with previous models, employing the H$_2$ formation rate on dust used in \citet{Gnedin2009}, \citet{Gnedin2011},  \citet{Christensen2012}, and discussed in our Section \ref{sec:h2formation}.
In Fig. \ref{fig:gnedin} we report the results with and without H$_2$ shielding together with our reference case where we do not have any shielding, and assume $C_{\rho}$ = 1, \mbox{$Z$ = 0.5 $Z_\odot$}, and  $z=3$ (Model Ia). An increase in metallicity and a decrease in redshift boost the formation of H$_2$ as already discussed in Section \ref{sec:metalparam}, together with a high clumping factor. Compared to our reference model (Model Ia), the transition to a fully molecular phase is reached around 10$^3$ cm$^{-3}$, which shifts to 10$^2$ cm$^{-3}$ once we also include the shielding. This is in agreement with previous 3D results \citep{Gnedin2009} which showed a fully molecular stage between 10$^2$--10$^3$~cm$^{-3}$ depending on the parameters employed (e.g. radiation strengths, dust-to-gas ratio, clumping factor etc.).
This test suggests that once we employ a weaker background, it mimics the effect of radiation attenuation, and the shielding has an impact on the final results, in this specific case shifting the HI-H$_2$ transition to lower densities by a factor of an order of magnitude.

To summarize, the results obtained by the suite of tests we performed in this work,  based on the models reported in Table \ref{tab:models}, suggest that:
\begin{itemize}
	\item the HI-H$_2$ transition strongly depends on the metallicity, the radiation background strength, and the clumping factor, three parameters which affect the H$_2$ formation rate on dust. However, in high-resolution studies, where the clumping factor is not needed, it is very important to employ an accurate treatment of the dust grains. We showed that employing a formation rate which depends on the grain size is more efficient in forming H$_2$.
	\item The non-equilibrium metal cooling is very important at low density ($n_{\mathrm{tot}} < 10$ cm$^{-3}$) and it is in general higher than what the equilibrium tables provide. We showed that this also depends on the physical conditions, i.e. density and temperature. 
	\item The shielding is important, but it should be coupled with a proper radiative transfer module, which can produce a reasonably attenuated radiation background taking into account the geometry and the photon history.
\end{itemize}
Overall, the models presented and tested in this work are appropriate to study the different gas phase transitions in galaxy simulations, and can follow the evolution of key observational tracers, such as CII.
This is the first chemical model for the ISM of galaxies publicly released through the package \kromes, and which can be employed in different hydrodynamical codes (e.g. \textsc{enzo}, \textsc{flash}, \textsc{ramses}, and \textsc{gasoline}), providing a high degree of flexibility. We discuss in the next Section some additional physical processes that could be added to the basic models we presented in the previous sections.

\subsection{Additional ingredients and final remarks}
To have a comprehensive chemical model for galaxies, additional ingredients can be considered besides the ones discussed in the previous Sections. Cosmic rays and X-ray primary and secondary ionisations can be included in the chemical network presented here together with the resulting Coulomb heating. While it has been shown by \citet{Richings2014} that secondary ionisation is not relevant, cosmic-ray ionisation becomes important in shielded regions as typical of molecular clouds. A list of important ionisation and dissociation reactions involving cosmic rays is reported in Table \ref{tab:cr} as taken from \textsc{kida} \citep{Wakelam2012}. The heating released into the gas by each of these reactions is assumed to be 20 eV. We note that there are detailed calculations on the heating released by cosmic-ray ionisation and a density dependent function has been provided for example in \citet{Glassgold2014}. The X-ray physics, on the contrary, is beyond the scope of this work but, as it is already part of the package \kromes, the users can incorporate it into the current framework if relevant.

Both for comparison with observational data and for a higher complexity the current network can be extended to include CO, as in the network by \citet{Glover2010} released with \kromes. However, we note that the aim of this work is to provide chemical models which can be easily (and without too much computational demand) embedded in hydrodynamical simulations and  we therefore decided to present low complexity models which are nevertheless  able to capture the most important physics necessary to follow the HII-HI-H$_2$ transition. In addition, it was shown by \citet{Glover2012} that the SFR might be insensitive to the molecular content, and that having H$_2$ or CO cooling is not a necessary condition to form stars. They showed that other processes, like the fine-structure emission by C$^+$ at low densities and the gas-grain energy transfer at high densities, are in fact very efficient coolants and that the shielding of the ISRF by dust is the main parameter affecting the star formation process. 

The models reported in this work not only include  state-of-the-art chemistry and microphysics,  but should be considered as an update to previous models (e.g. \citealp{Kruhmolz2008,Gnedin2009,Gnedin2011}) and also an extension and improvement, in particular for the accuracy of the different ingredients employed in \textsc{krome}.  The present framework is released with \krome and it is publicly accessible. It  represents one of the few attempts to share a common chemistry module that can be employed to assess a quantitative comparison between different hydrodynamical codes. We note that the  \textsc{agora} project \citep{Kim2014}, which aims to provide a robust code comparison of galaxy simulations, is mainly oriented to the study of the atomic phase, while here we present models which can be used for different gas phase transitions and most importantly can be extended and customized by the users.

\section{Conclusions}\label{sec:conclusions}
In this paper we have presented different chemical models (networks and microphysics) which can be employed in 3D simulations of galaxy formation and evolution, as well as in similar environments. We cover in fact a density range from 10$^{-8}$ to 10$^{4}$ cm$^{-3}$ and a large temperature range of 10--10$^9$ K. Note that the upper limit in densities is mainly dictated by the limit in the equilibrium cooling tables by \citet{Shen2010}. Our chemical network has been carefully checked to be valid in the whole range of temperatures. We included the most important cooling functions, heating processes, dust physics, and photochemical processes suitable for this kind of studies. These models have been presented within the framework of the astrochemistry package \kromes. An interface to some of the most used hydrodynamic codes is already released with the package (\textsc{Flash}, \textsc{Enzo}, and \textsc{Ramses}), but these models can be employed in any hydrodynamical code through an interface to \kromes, as we will show in a forthcoming paper where we use the smoothed-particle-hydrodynamic code \textsc{gasoline} to explore the gas phase transitions in an isolated galaxy simulation. The aim of the paper is to provide a flexible model where every single ingredient can be added/removed based on the applications and the computational demand of a given problem. 

Different one-zone collapse tests have been presented to study the effect of the most important ingredients on the thermal evolution of the gas and on the transitions between different phases. In particular, we have explored different  prescriptions for the H$_2$ formation on dust with focus on an accurate treatment of the dust physics. Employing a proper dust grain distribution and the accurate rate by \citet{Cazaux2009} boosts the formation of H$_2$, shifting the atomic-to-molecular transition to lower densities. However, we also noticed that under the presence of a strong ionising radiation background the gas remains ionised up to densities of 10 cm$^{-3}$ and the efficient photo-heating delays the formation of H$_2$. To elucidate this point we performed a series of tests where we varied the clumping factor (which boosts the H$_2$ formation rate on dust) and decreased the radiation background by a factor of 10. In both cases we found that the transition shifts to lower densities, an effect which is stronger in the case of a lower radiation background. It is then necessary to include a proper radiation attenuation in 3D simulations. We also tested the effect of introducing a simple recipe for the H$_2$ shielding, but since the thermal evolution is driven here by the photo-ionisation of hydrogen (and the consequent photo-heating), the effect of H$_2$ self-shielding turned out to be negligible. A radiative transfer algorithm is then more appropriate in order to accurately study the transitions between the different gas phases.

In the end, the main parameters discussed in this work need to be carefully checked based on the physical application and possibly on observational data. For instance, a combination of $C_\rho$ = 10, a higher rate for H$_2$ formation on dust, low-radiation background (e.g. at $z$ = 0), and shielding, leads to reasonable gas phase transitions, but this cannot be considered as the only applicable case.

We also discussed the effect of employing the non-equilibrium metal cooling below 10$^4$ K versus an equilibrium treatment. This is in general computationally more expensive, as we need to solve a linear system of equations ``on the fly" and add more species/reactions to our network, as we need to follow the non-equilibrium abundances of the various metals coolants. It turned out that by using the photo-ionisation equilibrium tables from \textsc{cloudy} below 10$^4$ K can lead to inaccurate results if, depending on the density, the system is far from  equilibrium. 

We show that the non-equilibrium metal cooling is in general higher than the photo-ionised equilibrium metal tables, and our equilibrium produces results which differ by about a factor of 2 compared to \textsc{cloudy}. We attribute these differences to the different metals, transitions, and chemical networks
employed. A non-equilibrium approach is in any case preferable for temperatures below 10$^4$ K, as also discussed, under different conditions and approaches, in previous papers \citep{Vasiliev2011,Oppenheimer2013,Richings2014}. The metal network is further important if we want to assess a comparison between theory and observations, for instance on the 158 $\mu$m CII emission band.


The chemical network and the \krome options files are publicly available at \url{http://bitbucket/tgrassi/krome} and ready to be used with the codes which are already interfaced with the package, but also via a custom interface to other framework codes. 
In addition, we can provide on request dust tables for different conditions when needed. 

\begin{acknowledgements}
We are very grateful to Sijing Shen for having provided the \textsc{cloudy} metal cooling tables and for fruitful discussions on this topic. SB, DRGS and RB thank the DFG for funding via the Schwerpunktprogram SPP 1573 ``Physics of the Interstellar Medium" under grants  \# BO 4113/1-2, SCHL 1964/1-2, and BA 3706/3-2. TG acknowledges the Centre for Star and Planet Formation funded by the Danish National Research Foundation. PRC acknowledges research funding by the Deutsche Forschungsgemeinschaft (DFG) Sonderforschungsbereiche (SFB) 963 and support by the Tomalla Foundation. We thank the referee for the useful comments.
\end{acknowledgements}

\bibliographystyle{aa}      
\bibliography{mybib_new} 

\appendix
\section{Dust temperature evolution}
\begin{figure}
	\begin{center}
	\includegraphics[scale=0.7]{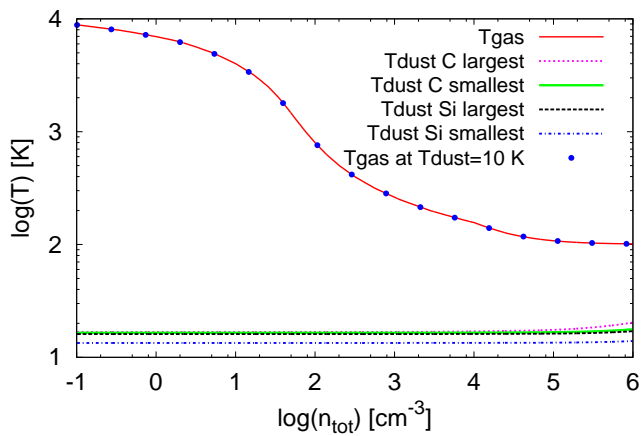}
	\caption{Thermal evolution for Model III and the dust temperature for the largest and smallest dust grains. Both silicates (Si) and carbonaceous (C) grains are reported. The dust temperature is obtained solving the thermal balance between the emitted and absorbed radiation and gas-grain thermal exchange (see Eq. \ref{eq:dustbalance} in the text).}\label{fig:dustrelax}
	\end{center}
\end{figure}

In this section we relax the assumption made on the dust temperature and let it evolve over time. In this case we keep the same grain distribution, i.e. 40 bins of carbonaceous and astronomical silicates, with a typical power-law ($dn/da\propto a^{-3.5}$) and, instead of employing the dust tables, we solve (by means of a root-finding procedure) the following thermal balance equation at run time:
\begin{equation}\label{eq:dustbalance}
	\Gamma_{\mathrm{em}} = \Gamma_{\mathrm{abs}} + \Lambda_{\mathrm{dust}}.
\end{equation}
The terms reported in the above equation depend on the grain size $a$, the radiation frequency $\nu$, and gas and dust temperatures, $T$ and $T_\mathrm{d}$. For details on Eq. \ref{eq:dustbalance} and on how every single term is defined see \citet{Grassi2014}. 
In Fig. \ref{fig:dustrelax} we report the thermal evolution of the gas together with the dust temperature for the different types of dust taking the largest and the smallest in size. We also plot the results obtained keeping the dust temperature constant at 10 K (same as in Fig. \ref{fig:model4}). As one can see from the figure, the dust temperature for all the grains is almost constant. It slightly increases only for the largest grains at high-density but the differences are not appreciable. The thermal evolution of the gas in the two cases is exactly the same. This confirms that the assumption to keep the dust temperature constant over the range of densities explored here is valid.

\section{Reaction rates employed in the present work}
In the following tables we list the reaction rates employed in the present work. These are publicly available with the \krome package under the folder \textsc{krome/networks}, named \textit{react\_galaxy}.
\begin{table*}
        \caption{List of reactions and rates included in our chemical network. Note that $T_e$ is the temperature in eV.}
        \begin{tabular}{@{}lllc}
                \hline\hline
                Reaction & Rate coefficient (cm$^3$ s$^{-1}$) & $T$ range & Ref.\\
                \hline
       1\dots\dots ~~H + e$^-$ $\rightarrow$ H$^+$ + 2e$^-$  & $k_1$ = exp[-32.71396786+13.5365560 ln $T_e$ & & 1 \\
        & $\phantom{k_{1}}$ - 5.73932875 (ln $T_e$)$^2$+1.56315498 (ln $T_e$)$^3$ & &  \\
& $\phantom{k_{1}}$ - 0.28770560 (ln $T_e$)$^4$+3.48255977 $\times$ 10$^{-2}$(ln $T_e$)$^5$ & & \\
        & $\phantom{k_{1}}$ - 2.63197617 $\times$ 10$^{-3}$(ln $T_e$)$^6$+1.11954395 $\times$ 10$^{-4}$(ln $T_e$)$^7$ & \\
        & $\phantom{k_{1}}$ - 2.03914985 $\times$ 10$^{-6}$(ln $T_e$)$^8$] & \\
        2\dots\dots ~~H$^+$ + e$^-$ $\rightarrow$ H  + $\gamma$ & $k_2$ = 3.92 $\times$ 10$^{-13}$ $T_e$ $^{-0.6353}$ & $T \le 5500$ K & 2 \\
        & $\phantom{k_{2}}$ = $\exp$[-28.61303380689232 & $T > 5500$ K & \\
& $\phantom{k_{2}}$- 7.241 125 657 826 851 $\times$ 10$^{-1}$ ln $T_e$\\
& $\phantom{k_{2}}$- 2.026 044 731 984 691 $\times$ 10$^{-2}$ (ln $T_e$)$^2$\\
& $\phantom{k_{2}}$- 2.380 861 877 349 834 $\times$ 10$^{-3}$ (ln $T_e$)$^3$\\
& $\phantom{k_{2}}$- 3.212 605 213 188 796 $\times$ 10$^{-4}$ (ln $T_e$)$^4$\\
& $\phantom{k_{2}}$- 1.421 502 914 054 107 $\times$ 10$^{-5}$ (ln $T_e$)$^5$\\
& $\phantom{k_{2}}$+ 4.989 108 920 299 510  $\times$ 10$^{-6}$ (ln $T_e$)$^6$\\
& $\phantom{k_{2}}$+ 5.755 614 137 575 750  $\times$ 10$^{-7}$ (ln $T_e$)$^7$\\
& $\phantom{k_{2}}$- 1.856 767 039 775 260  $\times$ 10$^{-8}$  (ln $T_e$)$^8$\\
& $\phantom{k_{2}}$- 3.071 135 243 196 590  $\times$ 10$^{-9}$  (ln $T_e$)$^9$]  \\
        3\dots\dots ~~He + e$^-$ $\rightarrow$ He$^+$ + 2e$^-$  & $k_3$ = $\exp$[-44.09864886 + 23.915 965 63 ln$T_e$ & & 1 \\
        & $\phantom{k_{3}}$- 10.753 230 2 (ln $T_e$)$^2$\\
        & $\phantom{k_{3}}$+ 3.058 038 75 (ln $T_e$)$^3$\\
        & $\phantom{k_{3}}$- 5.685 118 9 $\times$ 10$^{-1}$ (ln $T_e$)$^4$\\
        & $\phantom{k_{3}}$+ 6.795 391 23 $\times$ 10$^{-2}$ (ln $T_e$)$^5$\\
        & $\phantom{k_{3}}$- 5.009 056 10 $\times$ 10$^{-3}$ (ln $T_e$)$^6$\\
        & $\phantom{k_{3}}$+ 2.067 236 16 $\times$ 10$^{-4}$ (ln$T_e$)$^7$\\
        & $\phantom{k_{3}}$- 3.649 161 41 $\times$ 10$^{-6}$ (ln $T_e$)$^8$]  & \\
        4\dots\dots ~~He$^+$ + e$^-$ $\rightarrow$ He + $\gamma$ & $k_{4}$ =  3.92 $\times$ 10$^{-13}$ $T_e$ $^{-0.6353}$ &  $T_e \le 0.8$ eV & 3\\
        & $\phantom{k_{4}}$ =  3.92 $\times$ 10$^{-13}$ $T_e^{-0.6353}$ & $T_e > 0.8$ eV \\
        & $\phantom{k_{4}}$+ 1.54 $\times$ 10$^{-9}$ $T_e^{-1.5}$ [1.0 + 0.3 / $\exp$(8.099 328 789 667/$T_e$)]  \\
        & $\phantom{k_{4}}$ /[$\exp$(40.496 643 948 336 62/$T_e$)]\\
        5\dots\dots ~~He$^+$ + e$^-$ $\rightarrow$ He$^{++}$ + 2e$^-$ & $k_5$ = $\exp$[-68.710 409 902 120 01 + 43.933 476 326 35 ln$T_e$ &  & 4\\
        & $\phantom{k_{5}}$- 18.480 669 935 68 (ln $T_e$)$^2$ \\
        & $\phantom{k_{5}}$+ 4.701 626 486 759 002 (ln $T_e$)$^3$ \\
        & $\phantom{k_{5}}$- 7.692 466 334 492 $\times$ 10$^{-1}$ (ln $T_e$)$^4$\\
        & $\phantom{k_{5}}$+ 8.113 042 097 303 $\times$ 10$^{-2}$ (ln $T_e$)$^5$\\
                & $\phantom{k_{5}}$- 5.324 020 628 287 001 $\times$ 10$^{-3}$ (ln $T_e$)$^6$\\
        & $\phantom{k_{5}}$+ 1.975 705 312 221 $\times$ 10$^{-4}$ (ln $T_e$)$^7$ \\
        & $\phantom{k_{5}}$- 3.165581065665 $\times$ 10$^{-6}$ (ln $T_e$)$^8$] & \\
        \hline
\end{tabular}
\\1: \citet{Janev1987}, 2:  \citet{Abel97} fit by data from \citet{Ferland1992}, 3: \citet{cen92,Aldrovandi1973}\\
4: Aladdin database \url{https://www-amdis.iaea.org/ALADDIN/}, see \citet{Abel97}
\end{table*}\label{tab:rates}
         
\begin{table*}
        \contcaption{List of reactions and rates included in our chemical network.}
        \begin{tabular}{@{}lllc}
                \hline\hline
                Reaction & Rate coefficient (cm$^3$ s$^{-1}$) & $T$ range & Ref.\\
                \hline
           6\dots\dots ~~He$^{++}$ + e$^-$ $\rightarrow$ He$^+$ + $\gamma$ & $k_6$ = 
        1.891 $\times$ 10$^{-10}$(1.0+
       $\sqrt{T/9.37})^{0.2476}$ & & 5\\
       & $\phantom{k_{6}}$ (1.0+$\sqrt{T/(2.774\times 10^6)})^{1.7524}$  /($\sqrt{T/9.37})$& \\
7\dots\dots ~~H + e$^-$ $\rightarrow$ H$^-$ + $\gamma$ & $k_7$ = $1.4 \times 10^{-18} T^{0.928}\exp(-T/16200)$ & & 6\\
        8\dots\dots ~~H$^-$ + H $\rightarrow$ H$_2$ + e$^-$ & $k_8$ = $a_1(T^{a_2}+a_3 T^{a_4}+a_5T^{a_6})/(1.0+a_7T^{a_8}+a_9T^{a_{10}}+a_{11}T^{a_{12}})$ & & 7\\
        & $a_1 = 10^{-10}$\\
        & $a_2 = 9.8493 \times 10^{-2}$\\                                                                       
		& 	$a_3 = 3.2852 \times 10^{-1}$\\                                                                      
		& $a_4 = 5.5610 \times 10^{-1}$\\                                                                       
		& $a_5 = 2.7710 \times 10^{-7}$\\                                                                       
		& $a_6 = 2.1826$                           \\                                            
	 	& $a_7 = 6.1910 \times 10^{-3}$\\                                                                       
		& $a_8 = 1.0461$\\                                                                       
		& $a_9 = 8.9712 \times 10^{-11}$\\                                                                       
		& $a_{10} = 3.0424$\\                                                                      
		& $a_{11} = 3.2576 \times 10^{-14}$\\                                                                      
		& $a_{12} = 3.7741$\\
       9\dots\dots ~~ H + H$^+$ $\rightarrow$ H$_2^+$ + $\gamma$ & $k_9$ =  $2.10\times 10^{-20}(T/30.)^{-0.15}$ & $T < 30 $ K &  8\\
        & $\phantom{k_{9}}$ dex$[-18.20-3.194\log_{10}T+1.786(\log_{10} T)^{2}-0.2072(
        \log_{10}T)^3]$ & $T \geq 30$ K  \\  
        10\dots\dots ~~H$_2^+$ + H $\rightarrow$ H$_2$ + H$^+$ & $k_{10}$ = 6.0 $\times$ 10$^{-10}$ & & 9\\
        11\dots\dots ~~H$_2$ + H$^+$ $\rightarrow$ H$_2^+$ + H & $k_{11} $= 	dex$[\sum_{i=0}^{7} a_i [\log_{10}(T)]^i] $ & $10^2 \leq\ T \leq 10^8$ K & 10\\
        12\dots\dots ~~H$^-$ + e$^-$ $\rightarrow$ H + 2e$^-$  & $k_{12}$ = $\exp$[-18.018 493 342 73 & & 1 \\
        & $\phantom{k_{12}}$+ 2.360 852 208 681 ln$T_e$\\
        &$\phantom{k_{12}}$ - 2.827 443 061 704 $\times$ 10$^{-1}$ (ln $T_e$)$^2$\\
        & $\phantom{k_{12}}$+ 1.623 316 639 567 $\times$ 10$^{-2}$ (ln $T_e$)$^3$\\
        & $\phantom{k_{12}}$- 3.365 012 031 362 999 $\times$ 10$^{-2}$ (ln $T_e$)$^4$\\
        & $\phantom{k_{12}}$+ 1.178 329 782 711  $\times$ 10$^{-2}$ (ln $T_e$)$^5$\\
        & $\phantom{k_{12}}$- 1.656 194 699 504  $\times$ 10$^{-3}$ (ln $T_e$)$^6$\\
        & $\phantom{k_{12}}$+ 1.068 275 202 678  $\times$ 10$^{-4}$ (ln $T_e$)$^7$\\
        & $\phantom{k_{12}}$- 2.631 285 809 207  $\times$ 10$^{-6}$ (ln $T_e$)$^8$]& \\
  \hline       
\end{tabular}
\\5: \citet{Verner1996}, 6: \citet{DeJong1972}, 7: \citet{Kreckel2010}, 8: \citet{Coppola2011}, 9 \citet{Karpas1979}, 10: \citet{Savin2004} and \citet{Grassi2011}, see Table \ref{fitkrstic} for the coefficients.
\end{table*}

\begin{table*}
        \contcaption{List of reactions and rates included in our chemical network.}
        \begin{tabular}{@{}lllc}
                \hline\hline
                Reaction & Rate coefficient (cm$^3$ s$^{-1}$) & $T$ range & Ref.\\
                \hline
13\dots\dots ~~H$^-$ + H $\rightarrow$ 2H +  e$^-$ & $k_{13}$ = 2.56 $\times$ 10$^{-9}$ $T_e^{1.78186}$ & $T_e \le  0.04$ eV  & 11\\
        & $\phantom{k_{13}}$ =  $\exp$[-20.372 608 965 333 24 & $T_e >  0.04$ eV \\
        & $\phantom{k_{13}}$+ 1.139 449 335 841 631 ln $T_e$ \\
        & $\phantom{k_{13}}$- 1.421 013 521 554 148 $\times$ 10$^{-1}$ (ln $T_e$)$^2$\\
        & $\phantom{k_{13}}$+ 8.464 455 386 63 $\times$ 10$^{-3}$ (ln $T_e$)$^3$\\
        & $\phantom{k_{13}}$- 1.432 764 121 299 2 $\times$ 10$^{-3}$ (ln $T_e$)$^4$\\
        & $\phantom{k_{13}}$+2.012 250 284 791 $\times$ 10$^{-4}$ (ln $T_e$)$^5$\\
        & $\phantom{k_{13}}$+ 8.663 963 243 09 $\times$ 10$^{-5}$ (ln $T_e$)$^6$\\
        & $\phantom{k_{13}}$- 2.585 009 680 264 $\times$ 10$^{-5}$ (ln $T_e$)$^7$\\
        & $\phantom{k_{13}}$+ 2.455 501 197 039 2 $\times$ 10$^{-6}$ (ln $T_e$)$^8$\\
        & $\phantom{k_{13}}$- 8.068 382 461 18 $\times$ 10$^{-8}$ (ln $T_e$)$^9$]\\
        14\dots\dots ~~H$^-$ + H$^+$ $\rightarrow$ 2H + $\gamma$ & $k_{14}$ = $2.96 \times 10^{-6}/\sqrt{T}-1.73\times10^{-9}+2.50\times 10^{-10}\sqrt{T}-7.77\times10^{-13}T$ & & 12\\
        15\dots\dots ~~H$^-$ + H$^+$ $\rightarrow$ H$_2^+$ + e$^-$ & $k_{15}$ = 10$^{-8} \times T^{-0.4}$& & 13\\
       16\dots\dots ~~H$_2^+$ + e$^-$ $\rightarrow$ 2H + $\gamma$ & $ k_{16}$ = $1.0\times 10^{-8}$ & $T \le 617$ K & 14\\
 &$\phantom{k_{16}}$ = $1.32\times 10^{-6}T^{-0.76}$ & $T > 617$ K  \\
        17\dots\dots ~~H$_2^+$ + H$^-$ $\rightarrow$ H + H$_2$ & $k_{17}$ = 5.0 $\times$ 10$^{-7}$ (10$^2 \times T$)$^{-0.5}$& & 15 \\
        18\dots\dots ~~H$_2$ + H $\rightarrow$ 3H & $k_{18} = 6.67\times 10^{-12}\sqrt{T}\exp[-(1+63593/T)]$  & & 16\\
  19\dots\dots ~~H$_2$ + H$_2$ $\rightarrow$ H$_2$ + H + H & $k_{19} = 5.996\times 10^{-30}T^{4.1881}(1+6.761\times 10^{-6}T)^{-5.6881}\exp(-54657.4/T)$ & & 16\\
20\dots\dots ~~He$^+$ + H $\rightarrow$ He + H$^+$ & $k_{20} = 1.20\times10^{-15}(T/300)^{0.25}$ & & 17\\
21\dots\dots ~~He + H$^+$ $\rightarrow$ He$^+$ + H & $k_{21} = 1.26\times10^{-9}T^{-0.75}\exp(-1.275\times10^5/T)$ & $T \leq 10^4$ K & 18\\
& $\phantom{k_{21}}$ = $4\times10^{-37}T^{4.74}$ & $T > 10^4$ K \\
22\dots\dots ~~H$_2$ + e$^-$ $\rightarrow$ H + H + e$^-$ & $k_{22} = 4.38
\times10^{-10}T^{0.35}\exp(-102000/T)$ & & 19\\
23\dots\dots ~~H$_2$ + e$^-$ $\rightarrow$ H + H$^-$ & $k_{23} = 35.5\times T^{-2.28}
\exp(-46707/T)$ & & 20\\
24\dots\dots ~~H$_2$ + He $\rightarrow$ H + H + He & $k_{24} =$ dex$[-27.029 + 3.801\log_{10}(T)-29487/T]$ & & 21\\
25\dots\dots ~~H$_2$ + He$^+$ $\rightarrow$ He + H + H$^+$ & $k_{25} = 3.7\times 10^{-14}\exp(35.0/T)$ & & 22\\
26\dots\dots ~~H$_2$ + He$^+$ $\rightarrow$ H$_2^+$ + He & $k_{26} = 7.2\times 10^{-15}$ & & 22\\
27\dots\dots ~~H + H $\rightarrow$ H + H$^+$ + e$^-$ & $k_{27} = 1.2\times 10^{-17}T^{1.2}\exp(-157800/T)$ & & 24\\
28\dots\dots ~~H + He $\rightarrow$ He + H$^+$ + e$^-$ & $k_{28} = 1.75\times 10^{-17}T^{1.3}\exp(-157800/T)$ & & 24\\
29\dots\dots ~~H$^+$ + e$^-$ $\xrightarrow{dust}$ H & 
$k_{29} = 1.225\times 10^{-13}Z/Z_\odot [1.0+8.074\times 10^{-6}\psi^{1.378}$ & & 25 \\
 & $\phantom{k_{29}=} (1+5.087\times 10^2 T^{0.01586}\psi^{-0.4723-1.102\times 10^{-5}ln(T)}]^{-1}$ & & \\
30\dots\dots ~~He$^+$ + e$^-$ $\xrightarrow{dust}$ He & $k_{30} = 5.572\times 10^{-14}Z/Z_\odot [1.0+6.089\times 10^{-3}\psi^{1.1728}$ & & 25\\
 & $\phantom{k_{30}=}(1+4.331\times 10^2 T^{0.04845}\psi^{-0.8120-1.333\times 10^{-4}ln(T)}]^{-1}$ & &\\
 31\dots\dots ~~H + H $\xrightarrow{dust}$ H$_2$ & see text & & 26\\
 \hline       
\end{tabular}
11: \citet{Abel97} based on \citet{Janev1987}, 12: \citet{Stenrup2009}, 13: \citet{Poulaert1978}\\
14: \citet{Abel97},  fit from data by \citet{Schneider1994}
15: \citet{Dalgarno1987}, 16: \citet{Glover2008},\\ 17: \citet{Yoshida2006}, 18: \citet{Kimura1993}, 19: \citet{Mitchell1983} fit of data by \citet{Corrigan1965}\\ 20: \citet{Capitelli2007}, 21: \citet{Dove1987}, 22: \citet{Barlow1984},
24: \citet{Lenzuni1991}\\ 25: \citet{WeinDraine2001}, 26: \citet{Cazaux2009}, see also \citet{Grassi2014}
\end{table*}

\begin{table*}
\contcaption{List of reactions and rates included in our chemical network. Note that $\Psi$ has already been defined in Section \ref{sec:photoelectricH}.}
        \begin{tabular}{@{}lllc}
                \hline\hline
                Reaction & Rate coefficient (cm$^3$ s$^{-1}$) & $T$ range & Ref.\\
                \hline
32\dots\dots ~~$\cp + \me^- \rightarrow \mC  + \gamma$ &
$k_{32} = 4.67 \times 10^{-12}  \left(\frac{T}{300}\right)^{-0.6}$ & $T \le 7950 \: {\rm K}$ & 27 \\
& $\phantom{k_{32} } =1.23 \times 10^{-17}  \left(\frac{T}{300}\right)^{2.49} 
\exp \left(\frac{21845.6}{T} \right)$ & $ 7950 < T \le 21140 \: {\rm K}$ & \\
& $\phantom{k_{32}} = 9.62 \times 10^{-8} \left(\frac{T}{300}\right)^{-1.37} \exp \left(\frac{-115786.2}{T} \right)$ & $T > 21140 \: {\rm K}$ &\\
33\dots\dots ~~$\sip + \me^- \rightarrow \mSi + \gamma$ &
$k_{33} =  7.5 \times 10^{-12} \left(\frac{T}{300}\right)^{-0.55}$  & $T \le 2000 \: {\rm K}$ & 28 \\
 & $\phantom{k_{33}}= 4.86 \times 10^{-12} \left(\frac{T}{300}\right)^{-0.32}$ & $2000 < T \le 10^{4} \: {\rm K}$ & \\
& $\phantom{k_{33}}= 9.08 \times 10^{-14} \left(\frac{T}{300}\right)^{0.818}$ & $T > 10^{4} \: {\rm K}$ & \\
34\dots\dots ~~$\op + \me^-  \rightarrow \mO + \gamma$ &
$k_{34} = 1.30 \times 10^{-10} T^{-0.64}$ &  $T \le 400 \: {\rm K}$ & 29 \\
 & $\phantom{k_{34}} = 1.41 \times 10^{-10} T^{-0.66} + 7.4 \times 10^{-4}  T^{-1.5}$ & \\
 & $\phantom{k_{34}=} \mbox{} \times \exp \left(-\frac{175000}{T}\right) [1.0 + 0.062 \times 
\exp \left(-\frac{145000}{T}\right) ]$ & $T > 400 \: {\rm K}$ & \\
35\dots\dots ~~$\mC  + \me  \rightarrow \cp  + 2\me^- $ & 
$k_{35} = 6.85 \times 10^{-8} (0.193 + u)^{-1} u^{0.25} e^{-u}$ & $u = 11.26 / T_{\rm e}$ & 30 \\
36\dots\dots ~~$\mathrm{Si} + \me^-  \rightarrow \sip + 2\me^-$ & 
$k_{36} = 1.88 \times 10^{-7} (1.0 + u^{0.5}) (0.376 + u)^{-1} u^{0.25} e^{-u}$ & 
$ u = 8.2 / T_{\rm e}$ & 30 \\
37\dots\dots ~~$\mO  + \me^-  \rightarrow \op  + 2\me^-$ &
$k_{37} = 3.59 \times 10^{-8} (0.073 + u)^{-1} u^{0.34} e^{-u}$ & $u = 13.6 / T_{\rm e}$ & 30 \\
38\dots\dots ~~$\op  + \mH  \rightarrow \mO  + \Hp$ &
$ k_{38} = 4.99 \times 10^{-11} T^{0.405} +
7.54 \times 10^{-10} T^{-0.458} $ & & 31 \\
39\dots\dots ~~$\mO  + \Hp  \rightarrow \op  + \mH$ &
$k_{39} = [1.08 \times 10^{-11} T^{0.517}+ 4.00 \times 10^{-10} T^{0.00669}] \exp 
\left(-\frac{227}{T}\right) $ & & 32 \\
40\dots\dots ~~$\mO + \Hep \rightarrow \op + \He$ & 
$k_{40} = 4.991 \times 10^{-15} \left(\frac{T}{10000}\right)^{0.3794} 
 \expf{-}{T}{1121000}$ & & 33 \\
& $\phantom{k_{40} = } \mbox{} + 2.780 \times 10^{-15} 
\left(\frac{T}{10000}\right)^{-0.2163} \expf{}{T}{815800}$ & \\
41\dots\dots ~~$\mC  + \Hp  \rightarrow \cp  + \mH$ & $k_{41} = 3.9 \times 10^{-16} T^{0.213}$ & & 32 \\
42\dots\dots ~~$\cp  + \mH  \rightarrow \mC  + \Hp$ & $k_{42} = 6.08 \times 10^{-14} 
\left(\frac{T}{10000}\right)^{1.96} \expf{-}{170000}{T}$ & & 32 \\
43\dots\dots ~~$\mC + \Hep \rightarrow \cp + \He$ & 
$k_{43} = 8.58 \times 10^{-17}  T^{0.757}$ & $T \leq 200 \: {\rm K}$ & 34 \\
& $\phantom{k_{43}} = 3.25 \times 10^{-17} T^{0.968}$ & $200 < T \leq 2000 \: {\rm K}$ & \\
& $\phantom{k_{43}} = 2.77 \times 10^{-19} T^{1.597}$ & $T > 2000 \: {\rm K}$ & \\
44\dots\dots ~~$\mathrm{Si} + \Hp  \rightarrow \sip + \mH$ &
$k_{44} = 5.88 \times 10^{-13} T^{0.848}$ & $T \le 10^{4} \: {\rm K}$ & 35 \\
 & $\phantom{k_{44}} = 1.45 \times 10^{-13} T$ & $T > 10^{4} \: {\rm K}$ & \\
45\dots\dots ~~$\mathrm{Si} + \Hep \rightarrow \sip + \He$ & $k_{45} = 3.3 \times 10^{-9}$ & & 36 \\
46\dots\dots ~~$\cp  + \mSi \rightarrow \mC + \sip$ & $k_{46} = 2.1 \times 10^{-9}$ & & 36 \\
47\dots\dots ~~$\sip + \Hp \rightarrow \sipp + \mH$ & 
$k_{47} = 4.10 \times 10^{-10} \left( \frac{T}{10000} \right)^{0.24}$ & & 35 \\
 & $\phantom{k_{47} = } \mbox{} \times \left[1.0 + 3.17 \expf{}{T}{2.39 \times 10^{6}} \right] 
 \expf{-}{3.178}{T_{\rm e}}$ & & \\
48\dots\dots ~~$\sipp + \mH \rightarrow \sip + \Hp$ & $k_{48} = 1.23 \times 10^{-9} 
\left(\frac{T}{10000}\right)^{0.24}$ & & 35 \\ 
& $\phantom{k_{48} =} \mbox{} \times \left[1.0 + 3.17 \expf{}{T}{2.39 \times 10^{6}} \right] $ & & \\ 
49\dots\dots ~~$\sipp + \me^- \rightarrow \sip + \gamma$ & $k_{49} = 
1.75 \times 10^{-12} \left( \frac{T}{10000} \right)^{-0.6346}$ & & 37 \\ 
50\dots\dots ~~$\cp + \me^- \xrightarrow{dust} \mC$  &
$k_{50} = 4.558 \times 10^{-13} (Z/Z_{\odot}) [1.0 + 6.089 \times 10^{-3} \psi^{1.128} $ & & 25 \\
 & $\phantom{k_{50} =} (1.0 + 4.331 \times 10^{2} T^{0.04845} \psi^{-0.8120 - 
 1.333 \times 10^{-4}\ln T})]^{-1}$ & \\
51\dots\dots ~~$\op + \me^- \xrightarrow{dust} \mO$ & $k_{51} = \frac{1}{4} k_{29}$ & & 25,38 \\
52\dots\dots ~~$\sip + \me^- \xrightarrow{dust} \mSi$  &
$k_{52} = 2.166 \times 10^{-14} (Z/Z_{\odot}) [1.0 + 5.678 \times 10^{-8} \psi^{1.874} $ & & 25 \\
& $\phantom{k_{51}=} (1.0 + 4.375 \times 10^{4} T^{1.635\times10^{-6}} \psi^{-0.8964 - 
7.538 \times 10^{-5}\ln T})]^{-1}$ & \\
 \hline       
\end{tabular}
\\27: \citet{Nahar1997}, 28: \citet{Nahar2000}, 29: \citet{Nahar1999}, 30: \citet{Voronov1997},
31: \citet{Stancil1999}, 32: \citet{Stancil1998ApJ}
33: \citet{Zhao2004}, 34: \citet{Kimura1993},
35: \citet{Kingdon1996}, 36: \citet{LeTeuff2000}
37: \citet{Nahar1995,Nahar1996}, 38: \citet{Glover2007}
\end{table*}\label{tab:metals}

\begin{table*}
\caption{Cosmic rays related reactions}
        \begin{tabular}{@{}lll}
                \hline\hline
                Reaction & Rate coefficient (s$^{-1}$) \\
                \hline
53\dots\dots ~~$\mH + {\rm c.r.} \rightarrow \Hp + \me^-$ & 4.6$\times$10$^{-1}$  \\
54\dots\dots ~~$\He  + {\rm c.r.} \rightarrow \Hep + \me^-$ & 5.0$\times$10$^{-1}$ \\
55\dots\dots ~~$\mC  + {\rm c.r.} \rightarrow \cp + \me^-$ & 1.02$\times$10$^3$ \\
56\dots\dots ~~$\mO  + {\rm c.r.} \rightarrow \op + \me^-$ & 2.8 \\
57\dots\dots ~~$\mathrm{Si} + {\rm c.r.} \rightarrow \sip + \me^-$ & 4.23$\times$10$^3$  \\
58\dots\dots ~~H$_2$ + ${\rm c.r.} \rightarrow$ H + H & 0.1\\
59\dots\dots ~~H$_2$ + ${\rm c.r.} \rightarrow$ H$^+$ + H$^-$ & 3$\times$10$^{-4}$ \\
60\dots\dots ~~H$_2$ + ${\rm c.r.} \rightarrow$ H$_2^+$ + e$^-$ & 9.3$\times$10$^{-1}$\\
61\dots\dots ~~H$_2$ + ${\rm c.r.} \rightarrow$ H + H$^+$ + e$^-$ & 2.2$\times$10$^{-2}$\\
\hline
\end{tabular}
\\Note: all the rates are taken from the KIDA database  \citep{Wakelam2012}, and are expressed in terms of the cosmic rays ionisation of molecular hydrogen, $\zeta_{\mathrm{H_2}}$, which is an adjustable parameter and is approximately twice the value of $\zeta_{\mathrm{H}}$ \citep{Wakelam2012}. 
\end{table*}\label{tab:cr}

\begin{table*}
\caption{Fit coefficients for the charge transfer reaction $\mathrm
H_2+\mathrm H^+\to \mathrm H_2^++\mathrm H$. Coefficients are in the form $a(b)=a\times 10^b$.}
\vspace{1mm}
\begin{tabular*}{.35\textwidth}{c |r@{.}l r@{.}l}
\hline
$a_i$ &\multicolumn{2}{c}{$10^2\le T<10^5\ \mathrm K$} &\multicolumn{2}{c}{$10^5\le T \le 10^8\ \mathrm K$}\\
\hline
$a_0$ & $-1$&$9153214(+2)$          & $-8$&$8755774(+3)$            \\
$a_1$ & $ 4$&$0129114(+2)$          & $ 1$&$0081246(+4)$            \\
$a_2$ & $-3$&$7446991(+2)$          & $-4$&$8606622(+3)$            \\
$a_3$ & $ 1$&$9078410(+2)$          & $ 1$&$2889659(+3)$            \\
$a_4$ & $-5$&$7263467(+1)$          & $-2$&$0319575(+2)$            \\
$a_5$ & $ 1$&$0133210(+1)$          & $ 1$&$9057493(+1)$            \\
$a_6$ & $-9$&$8012853(-1)$          & $-9$&$8530668(-1)$            \\
$a_7$ & $ 4$&$0023414(-2)$          & $ 2$&$1675387(-2)$            \\
\hline
\end{tabular*}
\end{table*} \label{fitkrstic}

\end{document}